\documentclass[sigconf]{acmart}
\usepackage{graphicx}
\usepackage{subcaption}
\usepackage{enumitem}
\usepackage{xcolor}
\usepackage{xspace}
\usepackage{multirow}
\usepackage{algorithm}
\usepackage{amsmath}
\usepackage{stmaryrd} 
\usepackage{dsfont} 

\usepackage{listings}
\usepackage{soul} 
\usepackage{cleveref}

\usepackage[utf8]{inputenc}
\usepackage{xcolor}
\definecolor{softblue}{RGB}{65, 105, 225} 
\DeclareMathSymbol{\mlq}{\mathord}{operators}{``}
\DeclareMathSymbol{\mrq}{\mathord}{operators}{`'}

\definecolor{lightblue}{RGB}{245, 245, 220}
\definecolor{grey}{RGB}{169,169,169}
\definecolor{darkgreen}{RGB}{54, 156, 90}
\definecolor{lightblue}{RGB}{235, 247, 255}
\definecolor{niceblue}{RGB}{52, 164, 235}
\definecolor{nicered}{RGB}{219, 42, 48}
\definecolor{niceorange}{RGB}{224, 114, 54}
\definecolor{lightred}{RGB}{255, 227, 228}
\definecolor{lightorange}{RGB}{252, 235, 227}
\definecolor{orange1}{RGB}{214, 178, 131}
\definecolor{brown}{RGB}{163, 117, 57}
\definecolor{lightbrown}{RGB}{224, 163, 83}
\definecolor{commentcolor}{RGB}{101, 163, 178}
\definecolor{lightpurple}{RGB}{203, 195, 227}
\xdefinecolor{highlight_gold}{RGB}{204, 153, 0}

\definecolor{nicegreen}{RGB}{52, 207, 98}
\definecolor{lightgreen}{RGB}{223, 247, 230}
\usepackage[most]{tcolorbox}

\newtcbox{\highlight}{on line,
    arc=1pt, 
    colframe=red!20, 
    colback=red!20, 
    boxrule=0.0pt, 
    boxsep=0pt, 
    left=1pt, right=1pt, top=1pt, bottom=1pt 
}
\newtcbox{\highlightblue}{on line,
    arc=1pt, 
    colframe=blue!12, 
    colback=blue!12, 
    boxrule=0.0pt, 
    boxsep=0pt, 
    left=1pt, right=1pt, top=1pt, bottom=1pt 
}
\definecolor{codehighlight}{RGB}{237, 173, 12} 

\definecolor{highlightbg}{RGB}{237, 230, 12} 

\definecolor{marker-yellow}{RGB}{255, 255, 150} 
\definecolor{marker-green}{RGB}{200, 255, 200} 

\definecolor{highlight-yellow}{RGB}{255, 255, 150} 
\definecolor{highlight-blue}{RGB}{180, 210, 255} 

\newcommand{\projName}{HoarePrompt\xspace}
\newcommand{\benchName}{CoCoClaNeL\xspace}
\makeatletter
\def\mathcolor#1#{\@mathcolor{#1}}
\def\@mathcolor#1#2#3{%
  \protect\leavevmode
  \begingroup
    \color#1{#2}#3%
  \endgroup
}
\makeatother

\lstdefinestyle{largePython}{
    language=Python,
    basicstyle=\ttfamily\normalsize,
    keywordstyle=\bfseries\color{darkgrey},
    stringstyle=\color{darkgrey},
    commentstyle=\itshape\color{gray},
    showstringspaces=false,
    breaklines=true,
    frame=none,
    numbers=none,
     moredelim=[is][\color{red}]{@}{@},   
    moredelim=[is][\color{blue}]{!}{!}
}

\copyrightyear{2026}
\acmYear{2026}
\setcopyright{cc}
\setcctype{by}
\acmConference[ICSE '26]{2026 IEEE/ACM 48th International Conference on Software Engineering}{April 12--18, 2026}{Rio de Janeiro, Brazil}
\acmBooktitle{2026 IEEE/ACM 48th International Conference on Software Engineering (ICSE '26), April 12--18, 2026, Rio de Janeiro, Brazil}
\acmPrice{}
\acmDOI{10.1145/3744916.3773206}
\acmISBN{979-8-4007-2025-3/2026/04}

\begin{document}
\setlength{\parindent}{1em} 
\setlength{\parskip}{0pt} 
\title{\projName: Structural Reasoning About Program Correctness in Natural Language}

\author{Dimitrios Stamatios Bouras}
\email{2501112125@stu.pku.edu.cn}
\affiliation{%
  \institution{Key Laboratory of HCST (PKU), MoE, SCS, Peking University}
  \city{Beijing}
  \country{China}
}

\author{Yihan Dai}
\email{2501112020@stu.pku.edu.cn}
\affiliation{%
  \institution{Key Laboratory of HCST (PKU), MoE, SCS, Peking University}
  \city{Beijing}
  \country{China}
}

\author{Tairan Wang}
\email{tairan.wang.22@ucl.ac.uk}
\affiliation{%
  \institution{University College London}
  \city{London}
  \country{United Kingdom}
}

\author{Yingfei Xiong}
\email{xiongyf@pku.edu.cn}
\affiliation{%
  \institution{Key Laboratory of HCST (PKU), MoE, SCS, Peking University}
  \city{Beijing}
  \country{China}
}

\author{Sergey Mechtaev}
\email{mechtaev@pku.edu.cn}
\authornote{Sergey Mechtaev is the corresponding author.}
\affiliation{%
  \institution{Key Laboratory of HCST (PKU), MoE, SCS, Peking University}
  \city{Beijing}
  \country{China}
}

\authorsaddresses{} 

\newcommand{\jp}[1]{\textcolor{blue}{#1}}

\newcommand{\changes}[1]{%
#1\xspace
}

\authorsaddresses{} 

\begin{abstract}
While software requirements are often expressed in natural language, verifying the correctness of a program against such requirements is a hard and underexplored problem. Large language models (LLMs) are promising candidates for addressing this challenge, however, our experience shows that they are ineffective in this task, often failing to detect even straightforward bugs. To address this gap, we introduce \projName, a novel approach that adapts fundamental ideas from program verification to natural language artifacts. Inspired by the strongest postcondition calculus, \projName employs a systematic, step-by-step process in which an LLM generates natural language descriptions of reachable program states at various code points. To manage loops, we propose few-shot-driven k-induction, an adaptation of the k-induction method widely used in model checking. Once program states are described, \projName leverages the LLM to assess whether the program, annotated with these state descriptions, conforms to the natural language requirements. For evaluating the quality of classifiers of program correctness with respect to natural language requirements, we constructed \benchName, a challenging dataset of solutions to programming competition problems. Our experiments show that \projName improves the MCC by 61\% compared to directly using Zero-shot-CoT prompts for correctness classification. Furthermore, \projName outperforms a classifier that assesses correctness via LLM-based test generation by an MCC increase of 106\%. The inductive reasoning mechanism contributes a 26\% boost to MCC, underscoring its effectiveness in managing loops.
\end{abstract}

\maketitle

\section{Introduction}
\label{sec:introduction}

Assessing whether a program meets its requirements is a key part of software quality assurance. In practice, these requirements are often written in natural language, geared towards human-to-human communication or designed as prompts for LLMs. Verifying program correctness against such requirements is an inherently difficult and unexplored challenge. One approach is address it is to generate tests based on requirements using an LLM, and use these tests to check program correctness~\cite{chen2022codet}. More generally, natural language requirements can be translated in formal ones, and then used in conjunction with verification tools~\cite{endres2024can}. Despite the promise of these approaches, our experiments show that LLMs tend to generate incorrect or incomplete assertions when handling complex requirements, which reduces their accuracy. Another option is to directly employ LLMs to assess a program's conformance to requirements, relying on LLMs' ability to reason about both natural language and code~\cite{liu2024codemind}. Our experiments show that the state-of-the-art LLMs struggle with this task, often failing to identify even simple bugs. We attribute this limitation to the models' inability to reason about the intricacies of program semantics.

To address this gap, we introduce \projName, a program analysis approach for reasoning about program correctness w.r.t. natural language requirements that synergizes LLM reasoning techniques with ideas from program verification. In its core is the \emph{natural Hoare triple}, a variant of the traditional Hoare triple~\cite{hoare1969axiomatic} with pre- and postconditions expressed in natural language; for example, $$\mathcolor{softblue}{\{ \text{x is a full name} \}} \,\mathtt{x = x.split()[0]}\, \mathcolor{softblue}{\{ \text{x is a first name} \}}.$$ Inspired by the strongest postcondition calculus~\cite{dijkstra1976discipline}, \projName employs a step-by-step process to generate natural language descriptions of reachable program states at various points in the code. This is achieved by sampling \emph{natural strongest postconditions (NSP)} from $\mathcal{SP}$, our LLM-based natural Hoare triple completion model defining the conditional probability distribution $\mathcal{SP}(\mathcolor{softblue}{\mathrm{Post}}\,\mid\,\mathcolor{softblue}{\mathrm{Pre}}, C),$
where $\mathcolor{softblue}{\mathrm{Pre}}, \mathcolor{softblue}{\mathrm{Post}}$ are natural pre- and postconditions, $C$ is a program fragment. Inspired by methods to enhance LLM reasoning about program execution~\cite{ni2024next}, \projName uses these state descriptions to annotate a program, and then prompts an LLM to assess if the annotated program aligns with the given requirements.

\begin{figure*}[t]
  \begin{center}
    \includegraphics[width=\textwidth]{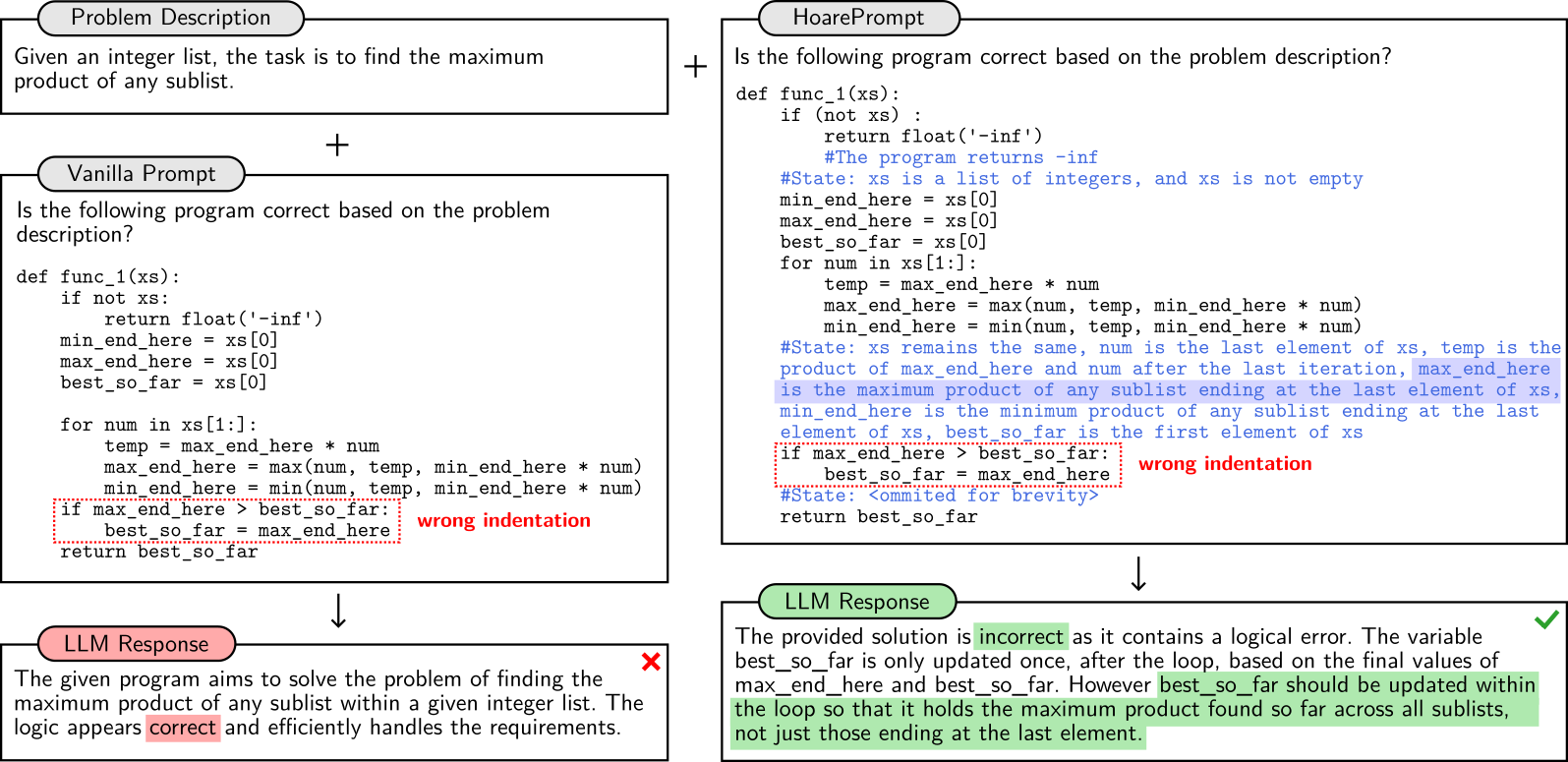}
  \end{center} 
  \caption{Annotating code with natural language descriptions of reachable program states, which are automatically inferred by \projName, enables an LLM to find a bug that is missed by the straightforward Vanilla prompt.\label{fig:motivating_example}}
  \vspace{-2mm}
\end{figure*}

\projName's key novelty in comparison with traditional program analysis and verification is in using natural language as the medium of reasoning about program semantics, as opposed to, e.g., first-order logic~\cite{filliatre2013why3} or monotonic functions over lattices~\cite{cousot1977abstract}. This enables a direct application of \projName to assessing program correctness w.r.t. natural language requirements without hard and time-consuming formal modeling. In comparison to previous techniques for reasoning about code with LLMs~\cite{ni2024next,liu2024codemind,jiang2024ledex}, \projName reasons about all reachable states simultaneously instead of individual concrete states. This enables \projName to operate statically, without executing the program and without reliance on provided failing tests. \changes{In comparison with directly prompting LLMs to analyze programs, \projName simplifies this task by decomposing it into small, independent subtasks, which are solved step-by-step in the spirit of chain-of-thought reasoning~\cite{wei2022chain}.}

Loops pose a fundamental challenge in program analysis and verification because they give rise to infinite execution paths~\cite{si2018learning}. Our experiments demonstrate that LLMs also struggle to summarize the semantics of complex loops, producing incorrect natural postconditions. To address this problem, we introduce a novel method, \emph{few-shot-driven k-induction}, synergizing few-shot learning~\cite{brown2020language} with the idea of k-induction~\cite{donaldson2011software} widely used in model checking. This method starts with iteratively unwinding a loop $k$ times to compute natural postconditions after each iteration. After that, it performs an induction inference step by prompting an LLM to compute the postcondition for the entire loop, using the computed natural Hoare triples for individual iterations as few-shot examples.

\begin{table}[t]
\centering
\caption{Success rates for detecting the bug in \Cref{fig:motivating_example} via prompts with and without state annotations, using four LLMs, across 60 responses for each, with the temperature 0.5.}
\vspace{-2mm}
\label{tab:motivating_example_results}
\begin{tabular}{l c c}
\toprule
\textbf{Model} & \textbf{Not Annotated} & \textbf{Annotated} \\
\midrule
Qwen2.5-7B          & 0\%   & 11.7\% \\
Qwen2.5-Coder-32B   & 15\%  & 85\%  \\
Qwen2.5-72B    & 35\% & 85\% \\
\changes{Llama3.1-70B}    & \changes{50\%} & \changes{100\%} \\
\bottomrule
\end{tabular}
\vspace{-3mm}
\end{table}

\begin{figure*}[t]
  \begin{center}
    \includegraphics[width=\textwidth]{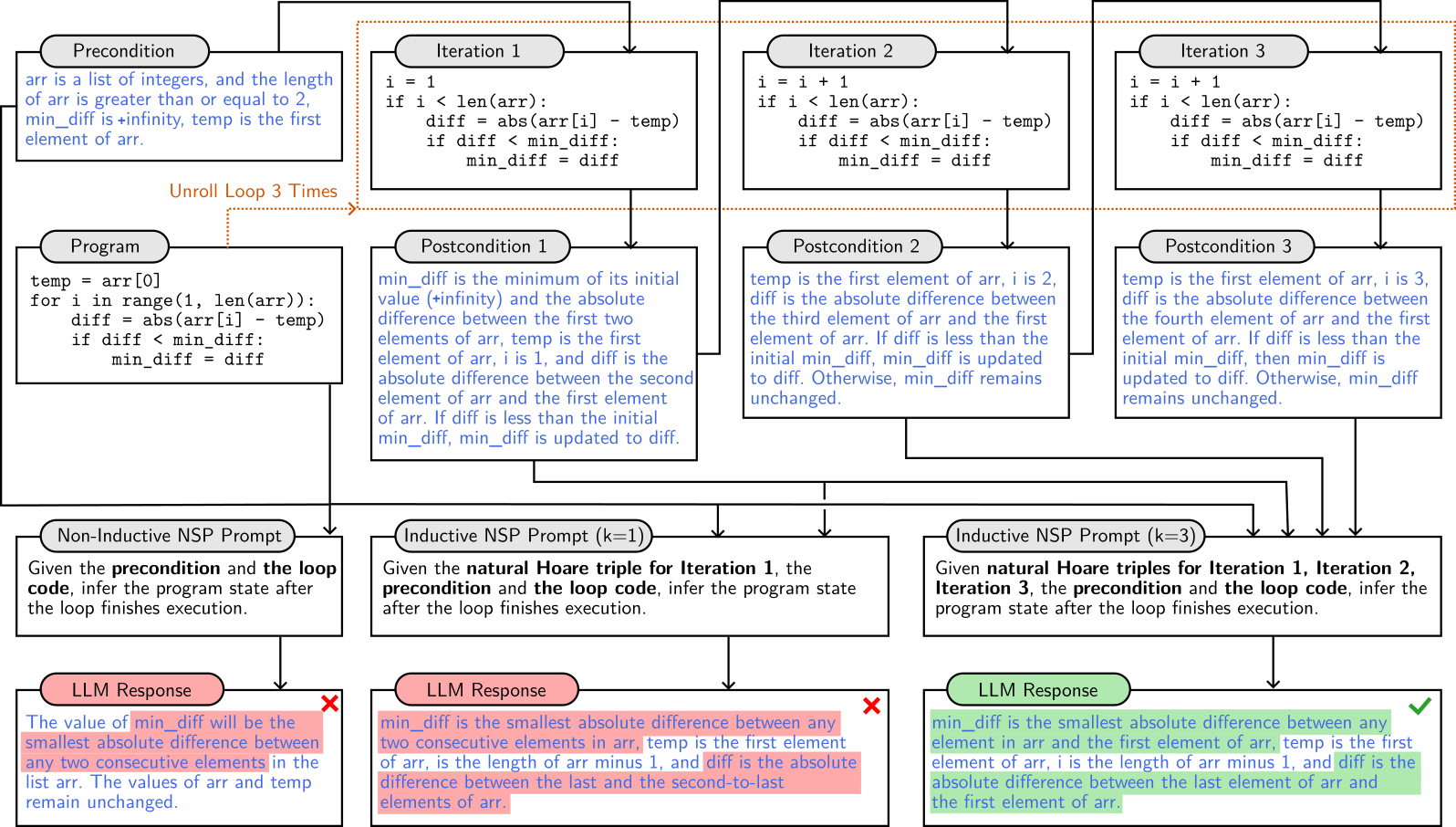}
  \end{center}
  \vspace{-2mm}  
  \caption{Few-shot-driven $k$-induction unrolls the loop three times, computes the natural strongest postconditions (NSP) for each iteration sequentially, and uses the resulting natural Hoare triples as few-shot examples for inductively inferring the postcondition for the entire loop, which enhances precision compared to non-inductive inference, or only a single unrolling.\label{fig:kinduction_example}}
\end{figure*} 
To evaluate \projName, we created \benchName, a dataset of program and natural language requirement pairs, where some programs meet the requirements and others do not due to human-introduced bugs. All programs and requirements were published since January 2024, after the training cut-off date of many SOTA LLMs to prevent data leakage. We formulated program correctness assessment as a binary classification problem. Our experiments show that \projName, at the expense of a higher token use, improves the Matthews Correlation Coefficient (MCC) by 61\% compared to directly using Zero-shot-CoT prompts. Furthermore, \projName outperforms a classifier that assesses correctness via LLM-generated tests by increasing the MCC by 106\%. The few-shot-driven $k$-induction mechanism contributes a 26\% boost to MCC. \changes{We also show that \projName improves code generation when used as a feedback mechanism, outperforming both direct generation (by 19.4\%) and test-based feedback (by 12.7\%).}

In summary, the paper makes the following contributions:
\begin{itemize}[nosep, topsep=0pt, partopsep=0pt, parsep=0pt, itemsep=0pt]
\item \projName, the first program analysis approach using natural language as the medium for reasoning about program semantics.
\item Few-shot-driven $k$-induction, the first approach that enhances LLM's ability to summarize loops.
\item \benchName, a challenging benchmark for program correctness classification.
\item An empirical study of classifying program correctness w.r.t. natural language requirements \changes{and guiding code generation}, demonstrating the effectiveness of \projName.
\end{itemize}
All code, scripts, and data, to reproduce this work are available at 
\changes{ Figshare: \href{https://figshare.com/s/5d39b388bd9ff2c7ffab}{https://figshare.com/s/5d39b388bd9ff2c7ffab}}. 
\vspace{-2mm}
\section{Motivating Examples}
\label{sec:motivating}

To motivate our approach, we first show that annotating a program with natural language descriptions of reachable program states enhances LLM's reasoning about program correctness, and then show how our approach precisely summarizes loops.
\vspace{-2mm}
\subsection{State Descriptions Enhance LLM Reasoning}
\label{sec:motivating_states}

Consider the following problem: ``given an integer list, the task is to find the maximum product of any sublist.'' \Cref{fig:motivating_example} (left) presents a flawed implementation of Kadane's algorithm to solve this problem. The issue lies in incorrect indentation, which results in \texttt{best\_so\_far} being updated only once, after the loop ends, instead of being updated during each iteration.

Despite the bug's simplicity, SOTA LLMs struggle to identify it consistently, and thus often wrongly classify the program as correct. Providing evidence that an LLM fails to solve a task is inherently challenging due to the model's non-deterministic behavior~\cite{ouyang2025empirical} and its sensitivity to prompt design~\cite{sclar2023quantifying}. To address these challenges, we made every effort to present objective evidence by crafting six distinct prompts: a straightforward (Vanilla) prompt, a Zero-shot-CoT prompt~\cite{kojima2022large}, an open-ended prompt~\cite{arora2022ask}, and combinations of these strategies, each containing the problem description and the code snippet. For each prompt, we generated 10 responses from four different LLMs, employing various temperature settings. More details about the experimental setup and results for this example can be found in the supplementary materials (\Cref{appendix:motivating_example}).

The results in \Cref{tab:motivating_example_results} demonstrate that LLMs, using the above six prompts, labelled collectively as ``Not Annotated'', detect the bug with only a small probability. An example of an incorrect response for the Vanilla prompt is presented in \Cref{fig:motivating_example} on the left.

\projName addresses this problem as follows. It first extracts the program's natural precondition by prompting an LLM to summarize the set of valid input values based on the problem description. In this case, the inferred precondition is $\mathcolor{softblue}{\{ \text{xs is a list of integers} \}}.$ Then, it iteratively queries its natural strongest postcondition (NSP) model to propagate reachable state descriptions along the program's control flow graph (CFG). For example, the Hoare Triplet for the first if statement:
\setlength{\jot}{0pt}
\begin{align*}
  &\mathcolor{softblue}{\{ \text{xs is a list of integers} \}}\\
  &\texttt{if (not xs):}\\
  &\quad\texttt{return float('-inf')}\\
  &\mathcolor{softblue}{\{ \text{xs is a list of integers, and xs is not empty} \}}
\end{align*}
\noindent capturing the fact that if an execution reaches the program point after the if statement, the list is guaranteed to be non-empty.

\projName continues to propagate the states until it reaches the end of the program. Loops, which present a particular challenge, are handled via inductive reasoning as illustrated in \Cref{sec:motivating_example_induction}.

Finally, \projName annotates the program with the reachable states at key program points (return statements, after top-level if statements and loops) as presented in \Cref{fig:motivating_example} on the right. Adding such annotations to the six prompts significantly increases LLM's success rate of correctness classification as shown in \Cref{tab:motivating_example_results}. \changes{This improvement stems from \projName’s compositional reasoning strategy. Since it reasons about individual blocks in isolation, the loop is interpreted based only on its local semantics. In this case, \projName accurately captures that the loop updates only consider sublists ending at the final element, with the resulting postcondition annotation enabling better error detection.}

\vspace{-3mm}
\subsection{Inductively Reasoning About Loops}
\label{sec:motivating_example_induction}

Reasoning about loops is challenging, because they induce infinite behaviours~\cite{si2018learning}, and our experience shows that LLMs also struggle to summarize them. To alleviate this problem, we take inspiration for the $k$-induction technique~\cite{donaldson2011software} widely used in model checking, which first checks that the property holds for all states in the first $k$ steps of the program's execution, and then proves that if a property holds for any k consecutive steps, then it holds for the next step.

To illustrate our adaptation of $k$-induction, which we refer to as \emph{few-shot-driven $k$-induction}, consider the program with a loop in \Cref{fig:kinduction_example}, labelled as ``Program''. Assume that its precondition is given in ``Precondition''. Inferring its natural strongest postcondition (NSP) using the straightforward ``Non-Inductive NSP Prompt'' with Qwen2.5-72B results in an incorrect natural postcondition mistakenly stating the \texttt{min\_diff} is the smallest absolute difference between any two consecutive elements in the list.

\projName unrolls the loop $k$ times, with the semantics of each iteration represented by code snippets in ``Iteration 1'', ..., ``Iteration k''. Then, it sequentially computes natural strongest postconditions for each iteration, using the postcondition of the previous iteration as the precondition of the next one (the first iteration uses the precondition of the entire loop). The resulting triples
\begin{align*}
  \mathcolor{softblue}{\{ \text{Precondition} \}} \,&\mathtt{Iteration\ 1}\, \mathcolor{softblue}{\{ \text{Postcondition 1} \}},\\
  &...\\
  \mathcolor{softblue}{\{ \text{Postcondition k-1} \}} \,&\mathtt{Iteration\ k}\, \mathcolor{softblue}{\{ \text{Postcondition k} \}}
\end{align*}
are used as few-shot examples to infer the postcondition of the entire loop in ``Inductive NSP Prompt (k=3)''. The intuition of this step is that LLMs generalize from few relevant examples~\cite{brown2020language}, and the examples above are relevant by construction. As shown in the corresponding LLM response, the enhanced prompt enabled the LLM to correctly infer that \texttt{min\_diff} is the smallest absolute difference between any  element in the list and the first element.

Interestingly, with k=1, the mistake persisted, as the model failed to generalize from a single example. Based on our pilot study (\Cref{sec:hyperparameters}), we chose $k=3$ as the best performing configuration.

\begin{figure}[t]
    \centering
    \includegraphics[width=\linewidth]{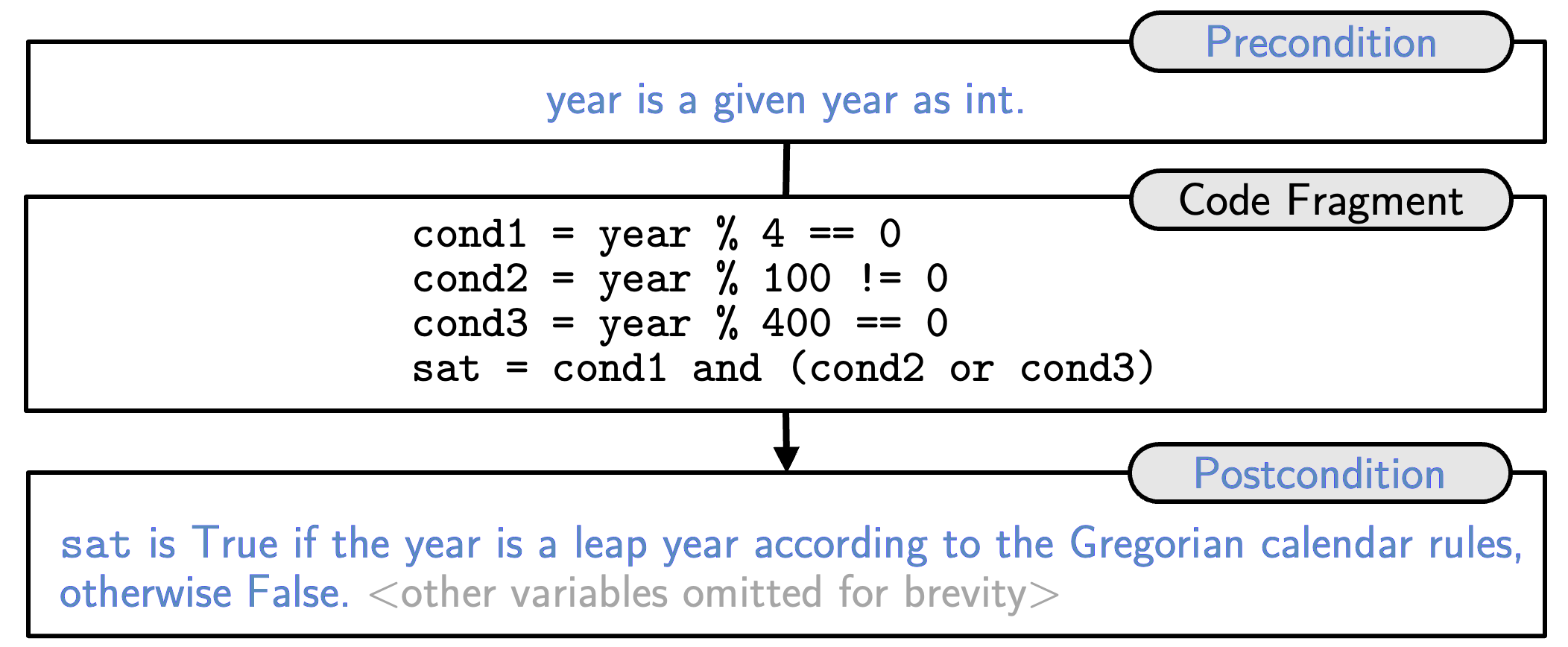}
    \vspace{-3mm}
    \caption{\projName succinctly describes program behavior in human-centric terms, which helps align formal semantics with informal requirements.
    \label{fig:motivating_expressive}}
    \vspace{-5mm}
\end{figure}
\vspace{-3mm}

\subsection{Expressive Power of Natural Language}
\label{sec:motivating_NL}
LLMs exhibit limitations in symbolic reasoning due to their reliance on natural language~\cite{xu2024faithful,can_llm_reason}. Consequently, \projName avoids formal logical inference, opting instead for natural language as the reasoning medium. While this approach does not provide formal correctness guarantees, it generates structured informal proofs, which, as demonstrated in \Cref{sec:evaluation}, improve the  LLMs' ability to effectively align program semantics with informal requirements.


 The method’s effectiveness stems from natural language’s expressivity. In \Cref{fig:motivating_expressive}, Qwen2.5-72B derives an NSP that succinctly describes \texttt{sat} in human terms, beyond formal semantics—an expressivity symbolic logic lacks. Further examples are in \Cref{appendix:NL}.



\vspace{-2mm}
\section{Background}
\label{sec:background}

Hoare logic~\cite{hoare1969axiomatic} is a formalism for reasoning about program correctness, at the core of which are specifications called \emph{Hoare triples} $\{\mathrm{Pre}\}\;C\;\{\mathrm{Post}\},$ meaning that the postcondition $\mathrm{Post}$ holds in any state reachable by executing the code $C$ from an initial state in which the precondition $\mathrm{Pre}$ holds, provided that the code terminates. Pre- and postconditions are predicates over program states, that are typically expressed in first-order logic~\cite{filliatre2013why3}. Strongest postcondition calculus~\cite{dijkstra1976discipline} automates verification using Hoare logic. The strongest postcondition (SP) of a program \( C \) w.r.t. a precondition \( \text{Pre} \), denoted as \( \text{sp}(\text{Pre}, C) \), is the most precise logical formula describing all possible states reachable after executing \( C \) from a state satisfying \( \text{Pre} \). Formally, $$\models \{\text{Pre}\}\;C\;\{\text{Post}\} \quad \text{iff} \quad \text{sp}(\text{Pre}, C) \implies \text{Post}.$$ The following algorithm for computing SP is defined compositionally for \textsc{While}~\cite{nielson2015principles}, a simple programming language that has only assignments ($x := e$), sequences ($S_1; S_2$), if statements, one loop instruction (while) and a single type (integer):
{\setlength{\abovedisplayskip}{2pt}
 \setlength{\belowdisplayskip}{2pt}
 \setlength{\jot}{1pt}
\begin{align*}
  \mathrm{sp}(\mathrm{Pre}, x := e) &= \mathrm{Pre}[x \mapsto e] \\
  \mathrm{sp}(\mathrm{Pre}, S_1; S_2) &= \mathrm{sp}(\mathrm{sp}(\mathrm{Pre}, S_1), S_2) \\
  \mathrm{sp}(\mathrm{Pre}, \texttt{if } b \texttt{ then } S_1 \texttt{ else } S_2) &=
      \mathrm{sp}(\mathrm{Pre \land b}, S_1) \vee
      \mathrm{sp}(\mathrm{Pre \land \neg b}, S_2) \\
  \mathrm{sp}(\mathrm{Pre}, \texttt{while } b \texttt{ do } S) &=
      (\mathrm{Pre} \land \neg b) \,\vee {}\\[-1mm]
  &\quad \mathrm{sp}(\mathrm{sp}(\mathrm{Pre \land b}, S),
      \texttt{while } b \texttt{ do } S)
\end{align*}}
\vspace{-2ex}

\noindent The last rule, which handles loops, is recursively defined, making it hard to compute.

The $k$-induction method~\cite{donaldson2011software} applies mathematical-induction principles to practical loop analysis. Let $P(s_i)$ denote the target property at loop state $s_i$ after $i$ iterations. The method has two components: the \emph{base case} (1), which checks that the property holds for the first $k$ iterations, and the \emph{inductive step} (2), which proves that any window of $k$ satisfying iterations implies the next one also satisfies it:
\[
\text{(1) }\bigwedge_{i=0}^{k-1} P(s_i)
\qquad\qquad
\text{(2) }\forall n\ge 0.\,\left( \bigwedge_{i=0}^{k-1} P(s_{n+i}) \right) \rightarrow P(s_{n+k})
\]

\section{\projName}
\label{sec:hoareprompt}

\begin{figure}
    \centering
    \includegraphics[width=0.75\linewidth]{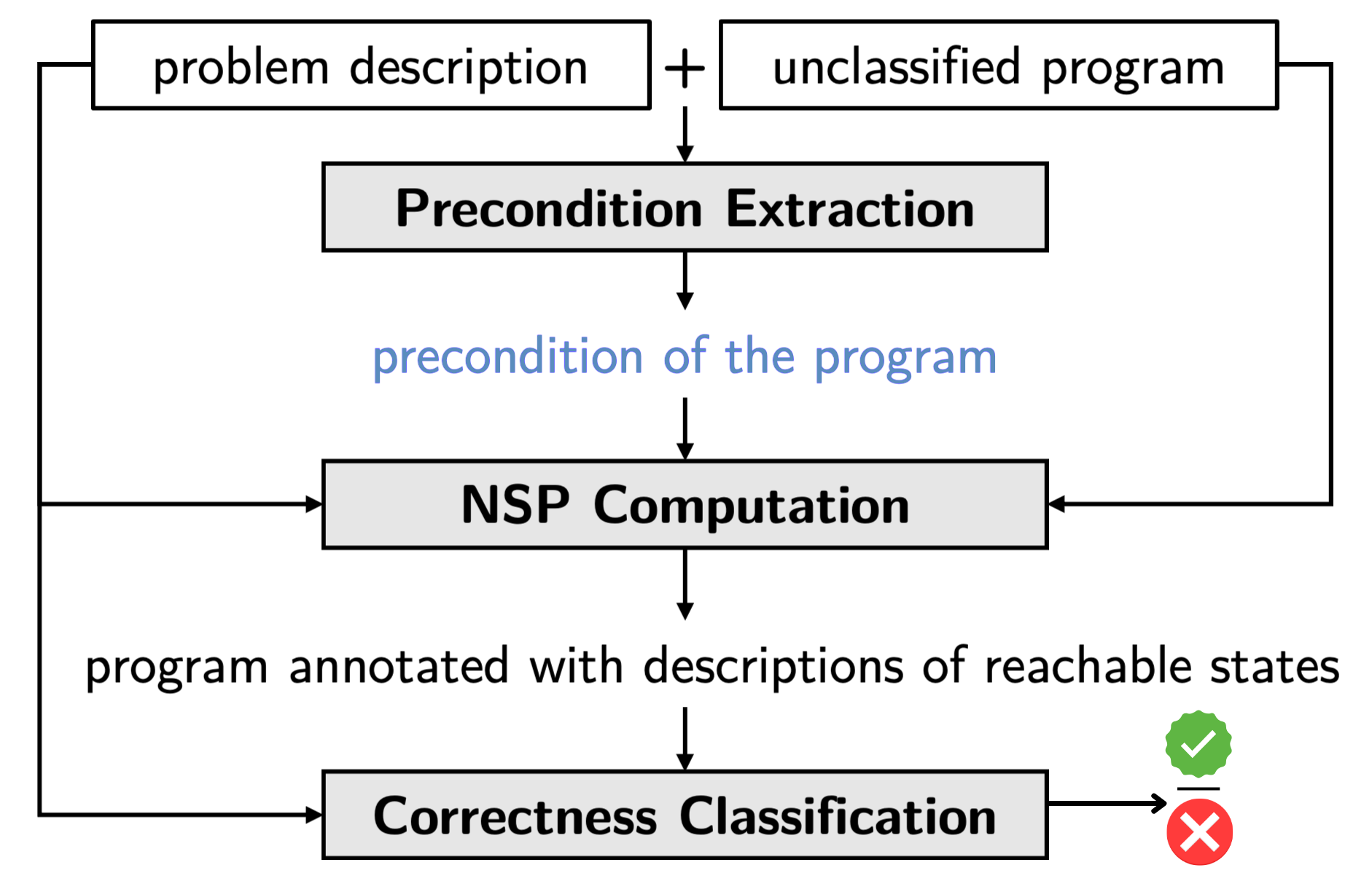}
    \vspace{-2mm}
    \caption{\projName's correctness classification workflow.\label{fig:workflow}}
    \vspace{-2mm}
\end{figure}

We introduce the \emph{natural Hoare triple}, a Hoare triple where pre- and postconditions are expressed in natural language (NL). To differentiate natural Hoare triples from traditional Hoare triples, we highlight sentences from NL with blue: $$\mathcolor{softblue}{\{\text{Pre}\}}  \, C \, \mathcolor{softblue}{\text{\{Post\}}}.$$

In analogy with the strongest postcondition calculus, we introduce the \emph{natural strongest postcondition (NSP)}. The traditional algorithm \( \text{sp}(\text{Pre}, C) \) cannot be applied in the context of NL. Instead, we create $\mathcal{SP}$, an LLM-based natural Hoare triple completion model defining a conditional probability distribution $$\mathcal{SP}(\mathcolor{softblue}{\mathrm{Post}} \mid\,\mathcolor{softblue}{\mathrm{Pre}}, \mathrm{Stmt}).$$ Let $\llbracket \cdot \rrbracket: \mathrm{NL} \to 2^S$ capture the human interpretation of NL state descriptions as a mapping from sentences to sets of program states $S$. Then, the goal to achieve by $\mathcal{SP}$ to ensure plausible reasoning can be measured using the following objective function:

\begin{definition}[NSP Objective Function] \label{def:objective} The objective function for $\mathcal{SP}$ is the cross-entropy loss between predicted postconditions and the traditional strongest postconditions computed over the interpretations of natural language preconditions: $$\mathcal{L} = -\sum_{\mathcolor{softblue}{\mathrm{Post}} \in \mathrm{NL}} \mathbb{I}(\llbracket \mathcolor{softblue}{\mathrm{Post}} \rrbracket \equiv \mathrm{sp}(\llbracket \mathcolor{softblue}{\mathrm{Pre}} \rrbracket, C)) \log \mathcal{SP}(\mathcolor{softblue}{\mathrm{Post}} \mid \mathcolor{softblue}{\mathrm{Pre}}, C),$$ where \( \mathbb{I}(\cdot) \) is the indicator function.
\end{definition}

\projName applies $\mathcal{SP}$ to classify program correctness w.r.t. natural language requirements using the workflow depicted in \Cref{fig:workflow}. First, it prompts an LLM to extract a precondition (a description of valid program inputs) from the requirements. Second, it traverses the program's control-flow graph (CFG), while sampling natural strongest postconditions from $\mathcal{SP}$, starting from the precondition, for each code segment. Finally, it annotates key program points with the inferred reachable state descriptions, and prompts an LLM to check whether the code meets the requirements.

\subsection{Precondition Extraction}

\projName prompts an LLM to extract a precondition from the given requirements in order to identify the set of valid program inputs. For example, for the problem description
\[
\text{``given $N$, write a function that finds the $N$th Fibonacci number''}
\]
and the function signature \texttt{def func(num)}, the LLM produces the precondition:
\vspace{-4mm}
\[
\mathcolor{softblue}{\{\text{num is a non\mbox{-}negative integer}\}}.
\]
We implement precondition extraction using few-shot learning (see supplementary materials, \Cref{pre_prompt,pre_multi_prompt}).


\begin{figure}[t]
    \centering
    \includegraphics[width=\linewidth]{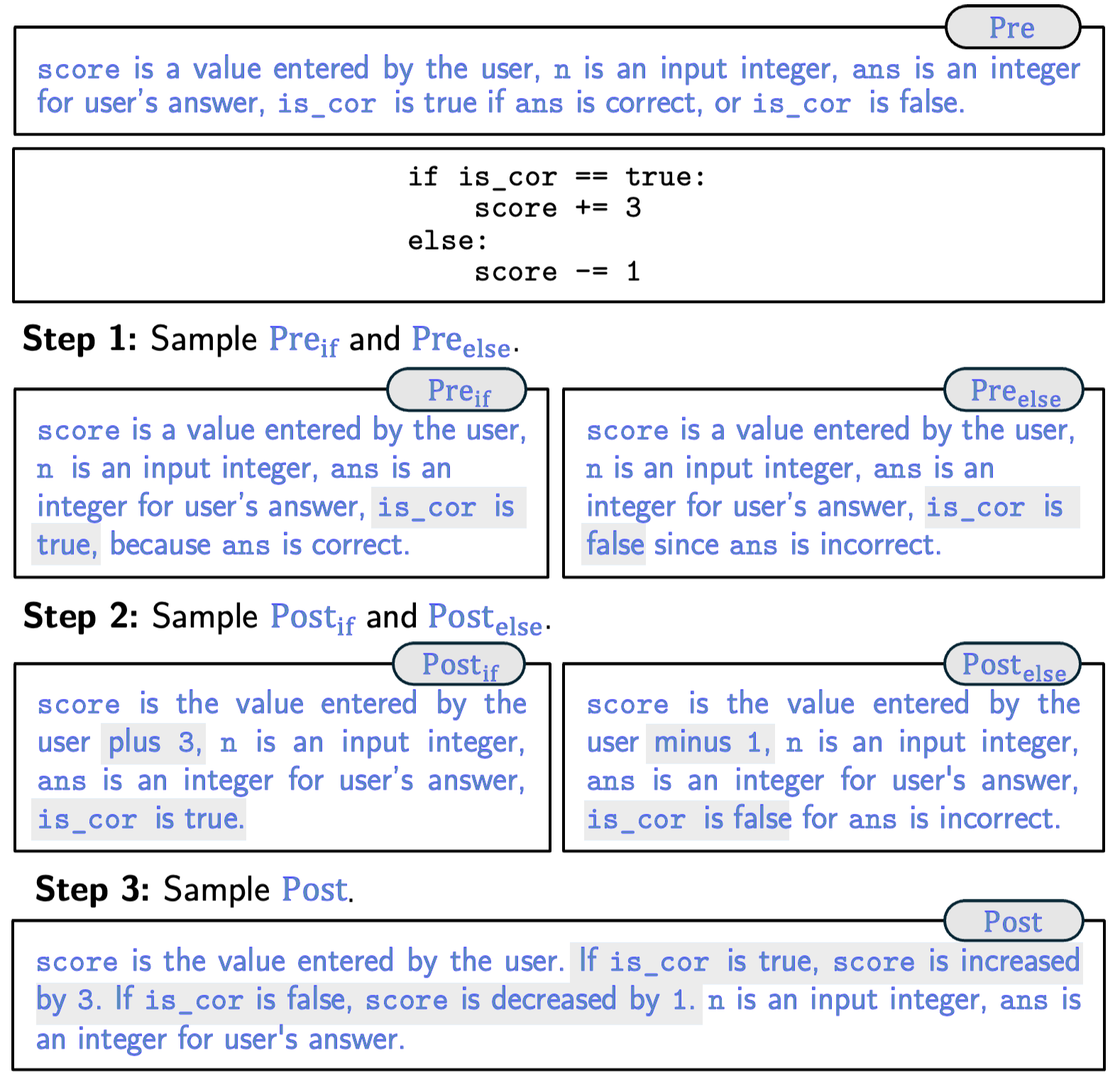}
    \caption{Computing NSP for an \texttt{if-else} statement.\label{fig:ifelse-example-label}}
    \vspace{-2mm}
\end{figure}

\subsection{NSP Computation}
\label{sec:nsp}

In principle, $\mathcal{SP}$ can be realized by fine-tuning an LLM to optimize the NSP objective function (\Cref{def:objective}). Due to budget constraints, we opted instead to adapt an LLM through prompt engineering and in-context learning. Specifically, we curated a pilot dataset, distinct from our evaluation set, and carefully designed prompts and few-shot examples for Python programs by eyeballing the LLM's output on this dataset.

\paragraph*{Simple Statements} Atomic operations such as assignments are processed by directly prompting an LLM to compute the postcondition. We use a dedicated prompt with few-shot examples, presented in supplementary materials \Cref{simple_statement_prompt}, that tasks the model to infer the updated state based on variable modifications while preserving relevant information from the precondition. Apart from that, we use dedicated prompts (in supplementary materials \Cref{return_prompt,print_prompt}) to capture effects of return and print commands, since program correctness often depends on them.

\paragraph*{Compound Statements} Similar to the traditional sp, $\mathcal{SP}$ is defined compositionally, which enables it to precisely reason about complex programs by breaking them down into simpler fragments. For example, given a sequence of statements $S_1; S_2$, $\mathcal{SP}$ is computed by first completing natural Hoare triple $\mathcolor{softblue}{\{\text{Pre}\}}S_1\mathcolor{softblue}{\{\text{Post}\}}$, and then its postcondition $\mathcolor{softblue}{\text{Post}}$ serves as the precondition for $S_2$. To reduce the number of LLM queries, we group a configurable number of atomic statements together using a dedicated prompt (in supplementary materials \Cref{compund statement prompt}). 


\paragraph*{If Statements} \Cref{fig:ifelse-example-label} shows how \texttt{if-else} statements are processed by \projName. Given a precondition $\mathcolor{softblue}{\{\text{Pre}\}}$ and a statement $\mathtt{if}\ b\!:S_1\ \mathtt{else}\!:\ S_2$, it first samples preconditions for the branches:
$$
\mathcolor{softblue}{\mathrm{Pre_{if}}}\sim \mathcal{SP}(\,\cdot\mid\,\mathcolor{softblue}{\mathrm{Pre}}, \mathtt{assert}\ b),\ \mathcolor{softblue}{\mathrm{Pre_{else}}}\sim \mathcal{SP}(\,\cdot\mid\,\mathcolor{softblue}{\mathrm{Pre}},  \mathtt{assert}\ \neg b)
$$
Then, it samples NSPs for each branch:
$$\mathcolor{softblue}{\mathrm{Post_{if}}}\sim \mathcal {SP}(\,\cdot\mid\,\mathcolor{softblue}{\mathrm{Pre_{if}}}, S_1),\
\mathcolor{softblue}{\mathrm{Post_{else}}}\sim \mathcal {SP}(\,\cdot\mid\,\mathcolor{softblue}{\mathrm{Pre_{else}}}, \mathrm{S_2})
$$
Finally, it uses a dedicated prompt to merge $\mathcolor{softblue}{\mathrm{Post_{if}}}$ and $\mathcolor{softblue}{\mathrm{Post_{else}}}$ into a single postcondition $\mathcolor{softblue}{\mathrm{Post}}$, which will hold after the loop. Prompts used for these operations are given in supplementary materials \Cref{if_pre_prompt,else_pre_prompt,if_else_prompt}.

\paragraph*{Loop Statements} \projName reasons about loops using our few-shot-driven $k$-induction method. For a while statement $\mathtt{while}\ b\!: S$ and a precondition $\mathcolor{softblue}{\mathrm{Pre}}$, it first samples a precondition for the loop body, and a postcondition for the first iterations: $$\mathcolor{softblue}{\mathrm{Pre_{while}}}\sim \mathcal{SP}(\,\cdot\mid\,\mathcolor{softblue}{\mathrm{Pre}}, \mathtt{assert}\ b),\ \mathcolor{softblue}{\mathrm{Post_1}}\sim \mathcal {SP}(\,\cdot\mid\,\mathcolor{softblue}{\mathrm{Pre_{while}}}, S).$$ Then, it continues to unwind the loop $k-1$ times, sequentially computing the following natural pre- and postconditions for each iteration:
$$\mathcolor{softblue}{\mathrm{Pre}_i}\sim \mathcal{SP}(\,\cdot\mid\,\mathcolor{softblue}{\mathrm{Post}_{i-1}}, \mathtt{assert}\ b),\ \mathcolor{softblue}{\mathrm{Post}_i}\sim \mathcal {SP}(\,\cdot\mid\,\mathcolor{softblue}{\mathrm{Pre}_i}, S),$$ where $i\in[2,3,...,k]$.

Finally, the NSP for the whole while statement is computed using our inductive prompt that includes the natural Hoare triples $$\Bigl\{\,\mathcolor{softblue}{\{\text{Pre}_i\}}  \, S \, \mathcolor{softblue}{\{\text{Post}_i\}}\,\Bigr\}_{i\in[1,2,...,k]}$$ as few-shot examples. This process is illustrated in \Cref{fig:kinduction_example} for a \texttt{for} statement, which is handled accordingly.

Many programming languages allow iterations over a collection, e.g., the {\tt for} loop in Python. 
The challenge of handling such iteration constructs is that they are implemented using implicit iterators, which store internal states updated at each iteration. Since iterators are not explicit variables, LLM tends to not include them when producing NSPs. Therefore, when dealing with iterators, we also explicitly model the state of the iterator.

Our modeling of Python's iterators involves maintaining the index of the next element of the iterator, which increments by 1 with each call to \texttt{next()}. Our natural language descriptions of program states refer to elements from the iterator as ``the \{index\}th elements of the iterator'', where ``\{index\}'' represents our modeled index, and the description of the iterator itself is always accompanied with the description of the current index representing its state. If an iterator is used across loops, its index also needs to be summarized via k-induction together with other parts of the state.

For a loop $\texttt{for}\ \texttt{item} \ \texttt{in I}$, to accurately reflect the additional changes of both the iterator and the index for  at the start of each iteration, the precondition for the i-th iteration is sampled as:
$$
\mathcolor{softblue}{\mathrm{Pre}_i}\sim \mathcal{SP}(\,\cdot\mid\,\mathcolor{softblue}{\mathrm{Post}_i}, \text{stmt}_\mathrm{iter}),
$$
where $i\in[1,2,...,k]$, $\mathcolor{softblue}{\mathrm{Post_0}}=\mathcolor{softblue}{\mathrm{Pre}}$, $\text{stmt}_\mathrm{iter}$ is 
\[
\begin{array}{l}
\texttt{try:}\\
\quad\texttt{item = next(I\_iter})\\
\quad\texttt{index += 1}\\
\texttt{except StopIteration:}\\
\quad\texttt{break}
\end{array}
\]
where \texttt{I\_iter} is an iterator obtained from \texttt{I} and \texttt{index} is initialized to \texttt{0}. For commonly used loops over lists or ranges of elements such as \texttt{range(0,n)}, we directly refer to the index of list without modeling the semantics of \texttt{next()} for simplicity.

The supplementary materials include our prompts for computing preconditions of the loop body (\Cref{while_first_prompt,for_first_prompt}), for handling \texttt{while} and \texttt{for} loops transitions (\Cref{while_next_prompt,for_next_prompt}), and for inductive inference (\Cref{total_while_prompt,total_for_prompt}).

\begin{figure}[t]
    \centering
    \includegraphics[width=\linewidth]{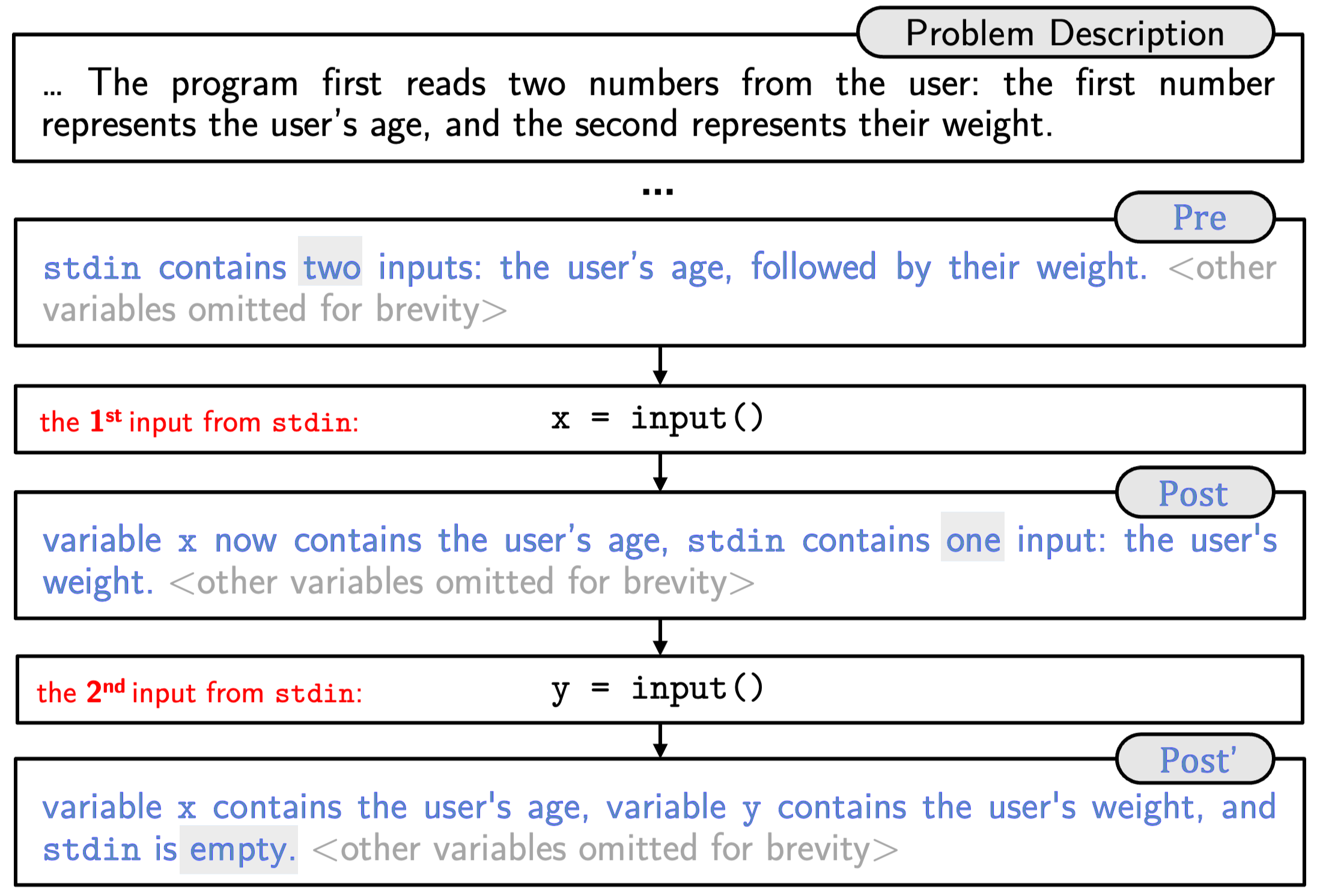}
    \vspace{-6mm}
    \caption{Computing NSP for the call of \texttt{input()} for \texttt{stdin}.\label{fig:stdin-example-label}}
    \vspace{-4mm}
\end{figure}

\paragraph{Exceptions Handling} Given $\mathcolor{softblue}{\mathrm{Pre}}$, and a \texttt{try-except} statement $$\text{try}\!:\alpha\ \text{except}\ E_1\!: \beta_1\ \text{except}\ E_2\!: \beta_2,\ldots\,\text{except}\ E_m\!: \beta_m,$$ we first sample the NSP for the \texttt{try}-block: $$\mathcolor{softblue}{\mathrm{Post_{try}}}\sim \mathcal{SP}(\,\cdot\mid\,\mathcolor{softblue}{\mathrm{Pre}}, \alpha),$$ and then the NSPs for different exception handling blocks: $$\mathcolor{softblue}{\mathrm{Post_i}}\sim \mathcal {SP}(\,\cdot\mid \mathcolor{softblue}{\text{``variables may take arbitrary values"}}, \beta_i).$$ 

Finally, we merge $\mathcolor{softblue}{\mathrm{Post_{try}}},\mathcolor{softblue}{\mathrm{Post_1}},\mathcolor{softblue}{\mathrm{Post_2}},\ldots,\mathcolor{softblue}{\mathrm{Post_m}}$ using a prompt provided in the supplementary materials (\Cref{try_except_prompt}). Note that, due to the complexity of handling nonlinear control flow, the NSP computation for exception handling blocks does not inherit any preconditions here, assuming that the state may be arbitrary.


\paragraph*{Input Stream} \Cref{fig:stdin-example-label} shows how data from the standard input (\texttt{stdin}) is processed by \projName. When programs read from \texttt{stdin}, the precondition extraction explicitly describes the expected structure and content of the \texttt{stdin} based on the given problem description. So, the precondition would clearly indicate that \texttt{stdin} contains these inputs in that specific order. Every time a postcondition for a statement reading $\texttt{stdin}$ is sampled as $$\mathcolor{softblue}{\mathrm{Post}}\sim \mathcal{SP}(\,\cdot\mid\,\mathcolor{softblue}{\text{Pre}}, \texttt{x\ = input()}),$$ the description of the stream state in $\mathcolor{softblue}{\mathrm{Post}}$ will be updated accordingly. After all data is read, the state of the stream is described as ``stdin is empty''. Note that this modeling applies only to non-interactive programs, where \texttt{stdin} is predetermined and does not depend on the execution of the program.

\subsection{Correctness Classification}
\label{sec:classification}

Let \( R \) be the natural language requirement, \( P \) be the program to be evaluated, \( D = \{(R_1, P_1, y_1), (R_2, P_2, y_2), \dots, (R_n, P_n, y_n)\} \) be a labeled dataset, where \( y_i \in \{ \textrm{Correct}, \textrm{Incorrect} \} \) represents whether program \( P_i \) satisfies requirement \( R_i \). The problem of classifying correctness w.r.t. natural language requirements can be formalized as the task of learning a function \( f \colon (R, P) \to \{\textrm{Correct}, \textrm{Incorrect}\} \).

In this work, we investigate several classification functions based on NSP, as well as alternative ideas.

{\textbf{\projName}:} Given $R$ and $P$, \projName first annotates key program points with descriptions of reachable states computed using $\mathcal{SP}$. Specifically, it adds annotations as comments after top-level if and loop statements. Apart from that, it adds descriptions of returned and printed values after return and print statements, as these values are often used to assess correctness. An example of an annotation program is given in \Cref{fig:motivating_example}. The annotated program is passed alongside $R$ to an LLM via our classification prompt. The supplementary materials contain the classification prompts for single- and multi-function programs (\Cref{correctness-single-func-prompt,correctness-multi-func-prompt}).

{\textbf{\projName (No Unroll)}: } We also implemented a variant of \projName that employs a non-compositional approach to computing the NSP for loops by prompting an LLM to infer the postcondition of the entire loop directly within a single LLM query. The prompts applied in this approach are given in the supplementary materials (\Cref{while_NO_UNROLL_PROMPT,for_NO_UNROLL_PROMPT}).

{\textbf{Vanilla}: } Given $R$ and $P$, it directly asks an LLM to classify program correctness using the prompt given in \Cref{vanilla-prompt}.

\textbf{\textbf{Zero-shot-CoT}:} Our Zero-shot-CoT~\cite{kojima2022large} prompt extends the Vanilla by triggering chain-of-thought reasoning~\cite{wei2022chain} by adding the ``think step-by-step'' instruction (\Cref{0-shot-COT-prompt}).

\textbf{\textbf{Tester}:} As an alternative to static reasoning, we also employed testing for classifying program correctness. Specifically, this method first generates tests based on $R$ with an LLM using the testing prompt adapted from AgentCoder~\cite{huang2023agentcoder}, executes $P$ on the generated tests, and classifies it as correct iff it passes the tests.

\section{Evaluation}
\label{sec:evaluation}

Our experiments address these research questions:
\begin{description}
\item[RQ1:] How does the classification performance of \projName compare to other techniques?
\item[RQ2:] How does few-shot-driven $k$-induction impact HoarePrompt’s effectiveness?
\item[RQ3:] How does the LLM token use of \projName compare to other techniques?  
\changes{\item[RQ4:] How effective is \projName as a feedback mechanism in a downstream code generation task?}
\end{description}

We selected four representative open-weight LLMs. Three come from the  Qwen~\cite{qwen} family: Qwen2.5-7B-Instruct, a popular smaller model, Qwen2.5-Coder-32B-Instruct, a model specialized for coding tasks, and Qwen2.5-72B-Instruct, demonstrating performance comparable to GPT-4o-mini~\cite{llm-stats}. \changes{Another one is from the Llama family: Llama3.1-70b-Instruct.} \changes{To measure classification quality, we adopted the Matthews Correlation Coefficient (MCC) as it considers all four components of the confusion matrix, and produces more reliable statistical results than F1 score~\cite{mcc}. Further information on MCC is available in supplementary materials (Appendix~\ref{appendix:mcc}.)} We also report other standard metrics such as Balanced Accuracy, True Positive Rate (TPR), False Negative Rate (FNR), and False Positive Rate (FPR). \changes{To ensure robustness against LLM non-determinism, we repeated each experiment multiple times. Following best practices~\cite{ouyang2025empirical,xiong2024llmsexpressuncertaintyempirical}, we calibrated the rerun count based on model confidence and consistency. This yielded 3 reruns per experiment. Details of this derivation are included in supplementary materials (Appendix~\ref{appendix:reruns})}. All experiments were conducted in a virtual machine with a AMD EPYC 7543 64-bit CPU, two cores and 16GB of RAM.

\begin{table}[t]
\centering
\caption{\changes{Effect of unroll bound $k$ on MCC and token usage (normalized over $k{=}0$).}}
\vspace{-2mm}
\label{tab:k-ablation}
\begin{tabular}{@{}lcc@{}}
\toprule
\changes{\textbf{Unroll Bound $k$}} & \changes{\textbf{MCC}}  & \changes{\textbf{Token Increase}} \\
\midrule
\changes{0 (no unroll)} &  \changes{0.211}  & \changes{1.00$\times$} \\
\changes{1} & \changes{0.228} & \changes{3.02$\times$} \\
\changes{3} & \changes{\textbf{0.262}}  & \changes{11.29$\times$} \\
\changes{5} & \changes{0.231} & \changes{24.53$\times$} \\
\bottomrule
\end{tabular}
\vspace{-2mm}
\end{table}

\begin{table*}[t]
    \centering
    \caption{Performance of classifiers of program correctness w.r.t. natural language requirements and their token use across four LLMs evaluated on \benchName. \projName significantly improves the MCC in comparison with baselines at the expense of a higher token use. \projName without loop unrolling represents a trade-off between classification quality and token cost. }
    \label{tab:results}
   \begin{tabular}{|c|l|p{0.9cm}p{1.2cm}p{0.9cm}p{0.9cm}p{0.9cm}p{0.9cm}|ccc|}

        \hline
        \multirow{2}{*}{\textbf{Model}} & \multirow{2}{*}{\textbf{Classifier}} & \multicolumn{6}{c|}{\textbf{Correctness Classification Quality Metrics}} & \multicolumn{3}{c|}{\textbf{Tokens (Millions)}} \\
        \cline{3-11}
        & & \textbf{MCC} & $\Delta\mathbf{MCC}$ & \textbf{BA} & \textbf{TPR} & \textbf{FNR} & \textbf{FPR} & \textbf{IT} & \textbf{OT} & \textbf{TT} \\
        \hline
       \multirow{5}{*}{\parbox{1.8cm}{\centering Qwen2.5-7B}}
         & Vanilla       & 0.055 & \textcolor{red}{  } & 0.521 & 0.193 & 0.807 & 0.151 & 1.8 & 0.8 & 2.6 \\
        & Zero-shot-CoT  & 0.085 & \textcolor{darkgreen}{+54.5\%}  & 0.543 & 0.563 & 0.437 & 0.478 & 2.1 & 1.3 & 3.4 \\
       
        & Tester             & 0.109 & \textcolor{darkgreen}{+98.18\%} & 0.519 & 0.051 & 0.949 & 0.012 & 1.7 & 3.8 & 5.5 \\
        \cline{2-11}
        & \textbf{HoarePrompt}         & \textbf{0.159} & \textcolor{darkgreen}{\textbf{+189.1\%}} & 0.577 & 0.700 & 0.300 & 0.546 & 205.1 (93.9) & 31.9 & 237.1 (125.9) \\
        & HoarePrompt no unroll & 0.119 & \textcolor{darkgreen}{+116.36\%} & 0.558 & 0.666 & 0.334 & 0.550 & 19.9 (12.3) & 3.4 & 23.4 (15.7) \\
        \hline
        \hline
        \multirow{5}{*}{\parbox{1.8cm}{\centering Qwen2.5-Coder-32B}}
        & Vanilla       & 0.137 & \textcolor{red}{ } & 0.553 & 0.233 & 0.767 & 0.128 & 1.9 & 0.9 & 2.8 \\
        & Zero-shot-CoT  & 0.123  & \textcolor{red}{-10.22\%} & 0.560 & 0.438 & 0.562 & 0.318 & 2.4 & 1.5 & 3.9 \\
        & Tester  & 0.096 &\textcolor{red}{-29.93\%}  & 0.518 & 0.055 & 0.945 & 0.019 & 1.9 & 3.3 & 5.2 \\
        \cline{2-11}
        & \textbf{HoarePrompt}         & \textbf{0.231} & \textcolor{darkgreen}{\textbf{+68.61\%}}  &0.616 & 0.593 & 0.407 & 0.362 & 186.5 (88.2) & 42.6 & 229.1 (130.8) \\
        & HoarePrompt no unroll & 0.158 &\textcolor{darkgreen}{+15.04\%}  & 0.579 & 0.571 & 0.429 & 0.412 & 19.2 (11.6) & 4.4 & 23.6 (15.9) \\
        \hline
        \hline
        \multirow{5}{*}{\parbox{1.8cm}{\centering Qwen2.5-72B}} 
        & Vanilla       & 0.169  & \textcolor{red}{ }  & 0.578 & 0.382 & 0.618 & 0.226 & 1.7 & 0.9 & 2.6 \\
        & Zero-shot-CoT  & 0.193 & \textcolor{darkgreen}{+14.20\%}  & 0.593 & 0.714 & 0.286 & 0.527 & 2.0 & 1.3 & 3.2 \\
        & Tester             & 0.131  &\textcolor{red}{-22.49\%}  & 0.522 & 0.052 & 0.948 & 0.007 & 1.9 & 3.8 & 5.6 \\
        \cline{2-11}
        & \textbf{HoarePrompt}         & \textbf{0.259} &  \textcolor{darkgreen}{\textbf{+53.25\%}} & 0.629 & 0.667 & 0.333 & 0.409 & 163.5 (78.7) & 32.5 & 196.0 (111.2) \\
        & HoarePrompt no unroll & 0.231 & \textcolor{darkgreen}{+36.69\%}  & 0.615 & 0.618 & 0.382 & 0.387 & 19.3 (11.8)  & 3.8 &  23.1 (15.6) \\
        
         \hline
        \multirow{5}{*}{\parbox{1.8cm}{\centering \changes{Llama3.1-70b}}} 
    & \changes{Vanilla}       & \changes{0.161}  & \textcolor{red}{ }  & \changes{0.579} &  \changes{0.671} & \changes{0.329} & \changes{0.513} & \changes{1.7} & \changes{0.01} & \changes{1.7} \\
    & \changes{Zero-shot-CoT}  & \changes{0.157} & \textcolor{red}{-2.48\%}  & \changes{0.575} & \changes{0.728} & \changes{0.329} & \changes{0.514} & \changes{1.8} & \changes{0.7} & \changes{2.5} \\
    & \changes{Tester}             & \changes{0.099}  & \textcolor{red}{-38.51\%}  & \changes{0.527} & \changes{0.109} & \changes{0.891} & \changes{0.055} & \changes{1.7} & \changes{2.4} & \changes{4.1} \\
    \cline{2-11}
    & \textbf{\changes{HoarePrompt}}         & \textbf{\changes{0.250}} &  \textcolor{darkgreen}{\textbf{+55.28\%}} & \changes{0.601} & \changes{0.895} & \changes{0.105} & \changes{0.692} & \changes{144.3 (106.8)} & \changes{27.3} & \changes{171.6 (134.1)} \\
    & \changes{HoarePrompt no unroll} & \changes{0.209} & \textcolor{darkgreen}{+29.81\%}  & \changes{0.588} & \changes{0.858} & \changes{0.142} & \changes{0.681} & \changes{12.0 (9.5)}  & \changes{1.7} &  \changes{13.7 (11.2)} \\
\hline

    \end{tabular}
   \caption*{\small \textbf{Abbreviations:} MCC = Matthews Correlation Coefficient, $\Delta\mathbf{MCC}$: Relative MCC Improvement over Vanilla , BA = Balanced Accuracy, TPR = True Positive Rate, FNR = False Negative Rate, FPR = False Positive Rate, IT = Input Tokens, OT = Output Tokens, TT = Total Tokens. Parentheses in IT and TT for HoarePrompt classifiers indicate token counts excluding few-shot learning examples, which can be eliminated via fine-tuning.}
   \vspace{-5mm}
\end{table*}

\subsection{\benchName Dataset}

Although there exist numerous datasets of natural language requirements~\cite{apps,mbpp}, some including incorrect implementations~\cite{tan2017codeflaws}, we could not use most of them due to potential data leakage, since memorization of correct solutions will significantly impact LLM reasoning~\cite{xie2024memorization,deng-etal-2024-unveiling}. Apart from that, we required non-trivial requirements and subtle bugs to thoroughly evaluate the ability of LLMs to reason about program correctness. We constructed a benchmark \benchName (\underline{Co}de \underline{Co}rrectness \underline{Cla}ssification w.r.t. \underline{N}atural \underline{L}anguage requirements). It consists of 645 problem-solution pairs from Codeforces programming contests that took place in the first half of 2024, with solutions labelled as correct or incorrect by the Codeforces' internal testing system. Totally, 322 of the solutions (49.9\%) are incorrect. All problems and solutions are released since January 2024, after the cut-off data of many SOTA LLMs, and the problems are challenging, have unambiguous descriptions, and solutions to these problems are rigorously tested to find subtle bugs~\cite{huang-etal-2024-competition}. To capture subtle and challenging coding errors, we specifically included incorrect submissions that immediately preceded correct ones, hypothesizing that these would likely contain nuanced mistakes. \changes{Further details about the dataset are provided in supplementary materials (\Cref{dataset_details})}.

\subsection{Hyperparameters}
\label{sec:hyperparameters}
\changes{We tuned the unrolling bound $k \in \{0,1,3,5\}$ using \texttt{Llama3.1-70B} on a 140-entry pilot set randomly sampled from the same source as \benchName, but disjoint from the 645 \benchName programs. As shown in \Cref{tab:k-ablation}, $k{=}3$ achieves the best MCC (0.262), balancing generalization and token cost. Higher $k$ increase token usage while degrading the quality. Notably, no matter the choice of $k$ \projName outperforms the baseline classifiers by at least 23.7\%.}

\changes{Another key design choice is how many and which annotations to include in the final prompt. We mark only major program points—after loops, branches, returns, prints, and try-catch blocks—to reduce prompt size while maintaining clarity. In pilots, full-state annotation improved MCC over Zero-shot-CoT by 19.7\%, whereas our sparse strategy achieved 62\%. We hypothesize that excessive detail dilutes focus, while key transitions aid reasoning.}

\subsection{RQ1: Classification Performance}

In the first research question, we examine whether structural reasoning with descriptions of reachable states helps \projName improve  LLMs' ability to evaluate program correctness compared to other techniques. By systematically  tracking program states through \projName, we hypothesize that structured reasoning will enhance correctness classification.


\Cref{tab:results} shows the results for each model. In comparison with LLM-based baselines, Vanilla and Zero-shot-CoT, \projName consistently demonstrated higher classification quality. On average, \projName improves the MCC by 61\% over Zero-shot-COT and 72\% over Vanilla. The impact of \projName is most pronounced in smaller models. The most significant improvement of MCC is in Qwen2.5-7B, being 87\% over Zero-shot-COT and 189\% over Vanilla. For Qwen2.5-72B, MCC improves by 34\%  and 53\% over the Zero-shot-CoT and Vanilla respectively. \changes{The results of the Llama model follow similar trends to those exhibited by the Qwen2.5 models, being especially close to Qwen2.5-72b, which is of a similar size.}

Similar improvement is observed across other classification metrics. \projName's TPR is 15.5\% higher than Zero-shot-COT and 107.1\% higher than Vanilla. \projName also reduces the FNR compared to Zero-shot-COT and Vanilla, The FPR for \projName (0.502), is also just slightly higher than Zero-shot-COT (0.459) but significantly higher than Vanilla (0.254).


\projName also significantly outperforms the \texttt{Tester} classifier, increasing MCC by 106\% on average. Notably, Tester achieves a very low average TPR of only 6.7\%, indicating frequent misclassifications of correct programs. This issue stems from test generation inaccuracies: LLMs often fail to produce correct output oracles, leading to false negatives. \changes{\Cref{tab:tester-stats} shows the breakdown of Tester behavior. On average, each program was evaluated with 13–28 test cases, but 15--25\% of examples had no passing test cases, and only 3--8\% had no failing test. Thus, the program was marked incorrect in over 90\% of cases, including the vast majority of correct programs, due to at least one failing test case. This affected all model sizes, showcasing a limitation of test-based correctness classification. A qualitative analysis of the misclassifications of \projName and Tester are presented in \Cref{qualitative}.}

\begin{table}[t]
\centering
\caption{\changes{Tester statistics across four LLMs. “Avg. Total” and “Avg. Passed” --- the average number of generated/passed tests per program. “0 Pass \%” and “0 Fail \%” --- the percentage of programs for which all tests failed or all tests passed.}}
\label{tab:tester-stats}
\begin{tabular}{@{}lcccc@{}}
\toprule
\changes{\textbf{Model}} & \changes{\textbf{Avg. Total}} & \changes{\textbf{Avg. Passed}} & \changes{\textbf{0 Pass \%}} & \changes{\textbf{0 Fail \%}} \\
\midrule
\changes{Qwen2.5-7B} & \changes{27.81} & \changes{9.99} & \changes{23.17\%} & \changes{3.32\%} \\
\changes{Qwen2.5-32B} & \changes{17.30} & \changes{7.86} & \changes{15.14\%} & \changes{3.63\%} \\
\changes{Qwen2.5-72B} & \changes{17.47} & \changes{6.50} & \changes{16.90\%} & \changes{3.13\%} \\
\changes{LLaMA3.1-70B} & \changes{12.72} & \changes{5.02} & \changes{25.34\%} & \changes{8.19\%} \\
\bottomrule
\end{tabular}
\vspace{-5mm}
\end{table}

\begin{tcolorbox}[colback=gray!5, colframe=black!20, arc=2pt, boxrule=0.3mm, breakable, sharp corners, left=2pt, right=2pt, top=2pt, bottom=2pt]
\textbf{RQ1:} \textit{\projName} consistently outperforms the baselines in terms of classification quality across all LLMs, achieving an average improvement of \textbf{61\%} in MCC over Zero-shot-COT, \textbf{72\%} over Vanilla and \textbf{106\%} over Tester.
\end{tcolorbox}

\subsection{RQ2: Impact of Inductive Reasoning}

Loops introduce complexity for reasoning about programs. \projName mitigates this complexity by inductively reasoning about loops using our novel few-shot-driven k-induction technique. This technique involves unrolling loops k times, with $k=3$ in our experiments, before generalizing the final state.

This question investigates whether loop unrolling  enhances correctness classification quality. To measure the impact of our few-shot-driven k-induction algorithm, we compare \projName's classification performance with that of \projName (no unroll) that summarizes loops within a single step.

As shown in \Cref{tab:results}, the few-shot-driven $k$-induction technique provides significant additional performance gains, improving the MCC by an average of 25.7\% over the non-inductive version (\textit{\projName} no unroll). Specifically, it improves MCC by 33.6\% in Qwen2.5-7B, 45.3\% in Qwen2.5-Coder-32B, 12.1\% in Qwen2.5-32B  \changes{and 19.6\% for LLama3.1-70b.} Even without unrolling, \projName demonstrates the second-best result among the studied classifiers.

\begin{tcolorbox}[colback=gray!5, colframe=black!20, arc=2pt, boxrule=0.3mm, breakable, sharp corners, left=2pt, right=2pt, top=2pt, bottom=2pt]
\textbf{RQ2:} Few-shot-driven k-induction enhances classification performance across all models, improving the MCC by an average of \textbf{25.7\%} over the non-inductive version of \projName.
\end{tcolorbox}


\subsection{RQ3: Token Usage}

Since \projName involves multiple LLM queries, it consumes more tokens than direct LLM prompts~\cite{liu2024groupdebate}. We quantified this increase and investigated the trade-off between token use and classification quality. Table~\ref{tab:results} provides the total token usage across classifiers and models for the 645 problem-solution pairs of the \benchName. For each classifier, we measured the input tokens (IT), the output tokens (OT), and total tokens (TT). As explained in \Cref{sec:nsp}, we chose to implement our natural strongest postcondition computation using prompt engineering with few-shot learning examples due to budget constraints. Tokens representing these hard-coded examples are repeatedly sent to an LLM during the algorithm iterations; in principle, they can be eliminated via fine-tuning. Because of that,  for the \projName variants, we also report the number of ``essential'' input and total token, in parentheses, that excludes hard-coded few-shot example tokens.

On average, \projName consumes 85.9 times more tokens than Vanilla (51 times when excluding non-essential tokens), 64.1 times more tokens than Zero-shot-CoT (38.6 times more when excluding non-essential tokens), and  40.8  times more tokens than Tester (24.6 times more, when excluding non-essential tokens).

A key reason for the increased token usage is our inductive reasoning algorithm (few-shot-driven $k$-induction) that performs loop unrolling. On average, it requires 10.9 times extra tokens (8.6 times extra when excluding non-essential tokens) than the non-inductive variant. Thus, the few-shot k-induction is more suitable for situations when classification accuracy is prioritized over cost. The non-inductive version, \projName no unroll, represents a practical trade-off by retaining a substantial performance boost (27.8\% better than the top performing approach, Zero-shot-COT) while consuming 90\% fewer tokens than full unrolling. Overall, it consumes only 6.4 times more tokens (4.5 times more, when excluding non-essential tokens) than Zero-shot-COT.

\begin{tcolorbox}[colback=gray!5, colframe=black!20, arc=2pt, boxrule=0.3mm, breakable, sharp corners, left=2pt, right=2pt, top=2pt, bottom=2pt]
\textbf{RQ3:} \projName's significant performance boost comes at the expense of an average token usage increase of 64 times (39 times when excluding non-essential tokens that can be eliminated via fine-tuning) compared to the most accurate baseline, Zero-shot-COT. The version of \projName without unrolling represents a practical trade-off between classification quality and cost, demonstrating the second-best performance, while consuming 90\% fewer tokens than the main variant.
\end{tcolorbox}

\subsection{RQ4: Code generation}
\label{sec:codegen}

\changes{To assess \projName as a feedback mechanism in the context of code generation, we adapted the AgentCoder~\cite{huang2023agentcoder} pipeline and implemented two variants: one using test-based feedback (as in the original), and one replacing test failures with correctness judgments from \projName. The third baseline generates a final solution in a single step without refinement. We evaluated all three methods on 100 problems randomly sampled from \benchName. We excluded problems with multiple valid outputs --- cases where different programs produce acceptable but non-identical outputs --- since correctness becomes ambiguous and not reliably testable~\cite{quan2025codeelobenchmarkingcompetitionlevelcode}.}

\changes{For our main experiments, we used \texttt{Llama3.1-70b-instruct}. Due to the complexity of \benchName and to evaluate generalization on more capable models, we additionally employed the stronger \texttt{GPT-4.1} model.} \changes{Results for both models are shown in \Cref{tab:agentcoder_results_combined}.}

\begin{table}[t]
\centering
\caption{\changes{Code generation results on 100 problems from \benchName. 
\textbf{Correct\%}: percentage of programs passing all hidden tests; 
\textbf{ATP}: average fraction of test cases passed.}}
\label{tab:agentcoder_results_combined}
\begin{tabular}{lcccc}
\toprule
\multirow{2}{*}{\changes{\textbf{Method}}} & 
\multicolumn{2}{c}{\changes{\textbf{Llama3.1}}} & 
\multicolumn{2}{c}{\changes{\textbf{GPT-4.1}}} \\
\cmidrule(lr){2-3} \cmidrule(lr){4-5}
 & \changes{\textbf{Correct\%}} & \changes{\textbf{ATP}} 
 & \changes{\textbf{Correct\%}} & \changes{\textbf{ATP}} \\
\midrule
\changes{\projName} & \changes{\textbf{33\%}} & \changes{\textbf{57.98\%}} & \changes{\textbf{47\%}} & \changes{\textbf{69.55\%}}\\
\changes{Testing}   & \changes{29\%} & \changes{54.89\%} & \changes{42\%} & \changes{69.53\%}\\
\changes{Baseline}  & \changes{30\%} & \changes{56.34\%} & \changes{37\%} & \changes{62.03\%}\\
\bottomrule
\end{tabular}
\end{table}

\changes{With Llama3.1-70B, the testing-based variant occasionally regressed due to incorrect oracles, reducing both fully correct programs and test pass rates compared to the baseline. In contrast, \projName consistently improved both metrics, highlighting its robustness. The supplementary materials include a regression example (\Cref{ex_agent_coder_test}) and \projName's improvement (\Cref{ex_hoareprompt_beautiful_pairs}). With the stronger GPT-4.1, the test-based variant improves over the baseline --- suggesting that test-driven refinement benefits from stronger models. However, \projName still outperforms both, achieving the highest correct rate (47\%) and best ATP (69.55\%).}


\begin{tcolorbox}[colback=gray!5, colframe=black!20, arc=2pt, boxrule=0.3mm, breakable, sharp corners, left=2pt, right=2pt, top=2pt, bottom=2pt]
\changes{\textbf{RQ4:} \projName is effective as a feedback mechanism for iterative code generation, increasing the number of solved problems by 12--14\% compared to using generated tests as feedback.}
\end{tcolorbox}

\subsection{Qualitative Error Analysis}
\label{qualitative}

\begin{figure*}[t]
  \centering
  \includegraphics[width=\textwidth]{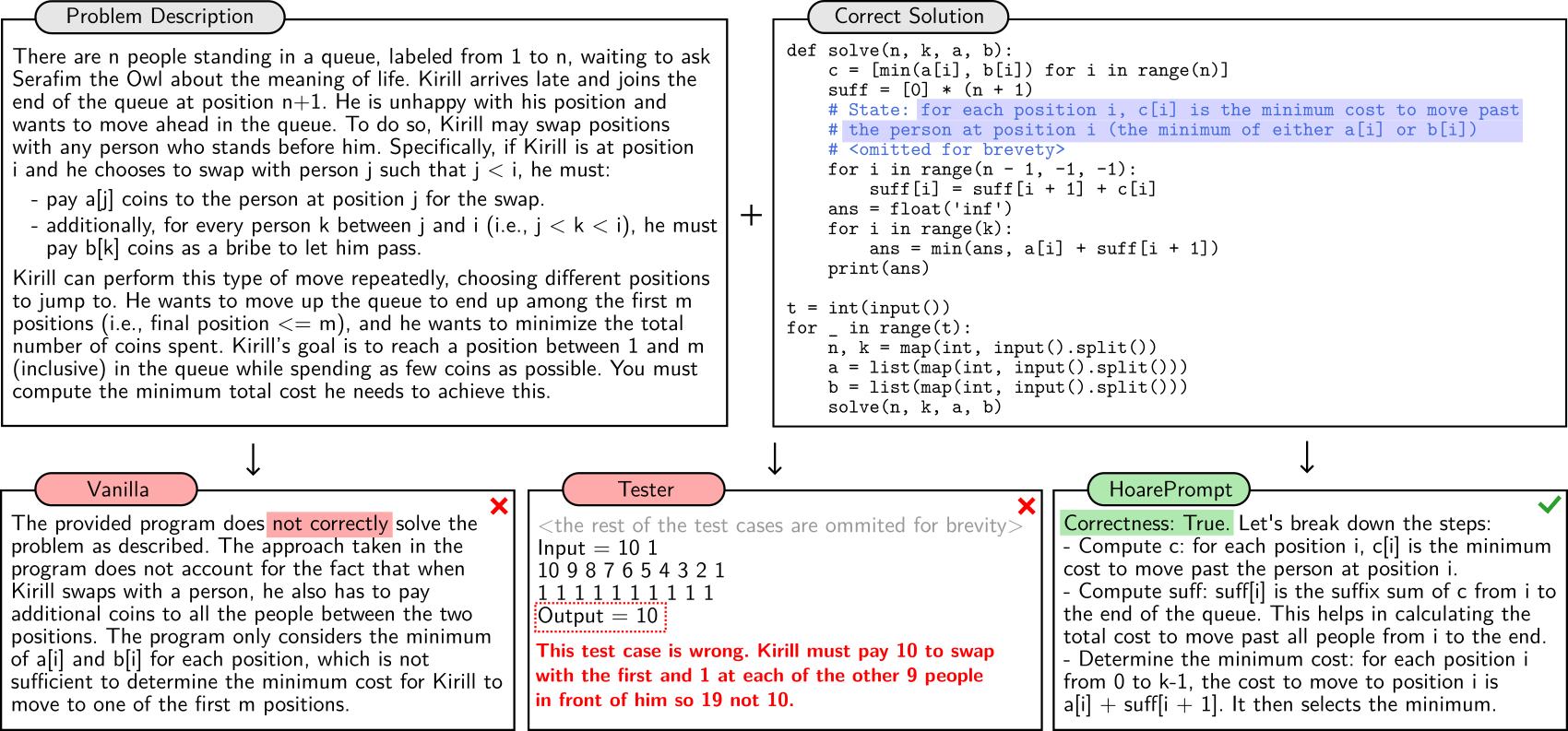}
  \caption{\changes{A \benchName problem with a correct solution. \textsc{Vanilla} misclassifies by misreading the algorithm; \textsc{Tester} fails by predicting wrong outputs; \projName’s state annotations enable a correct judgment.}\label{fig:tester_example}}
  \vspace{-2mm}
\end{figure*}

\changes{We manually audited misclassifications to understand failure modes of \projName and Tester.}First for \projName:
\begin{itemize}[leftmargin=*,nosep]
  \item \changes{\textbf{Missing Edge Case:}} \changes{The NSP chain omits critical boundary conditions, leading to overly optimistic judgments.}

  \item \changes{\textbf{Oversimplified Annotations:}} \changes{Annotations blur fine-grained state differences, producing misleading correctness cues.}

  \item \changes{\textbf{Incorrect Loop Generalization:}} \changes{Loop effects are overgeneralized, resulting in faulty postconditions.}

  \item \changes{\textbf{Semantic Trust Bias:}} \changes{When code structure resembles the description, the LLM sometimes ignores contradictory annotations and falsely assumes correctness.}

  \item \changes{\textbf{Interprocedural Failures:}} \changes{Reasoning errors occur across function boundaries, corrupting inference consistency.}

  \item \changes{\textbf{Variable Scope or Shadowing Errors:}} \changes{In loop-heavy code, reused indices or scope bleed distort postconditions.}

  \item \changes{\textbf{Description–Code Misalignment (rare):}} \changes{For correct programs using alternative logic patterns (e.g., DP vs.\ greedy), annotations fail to reflect intent.}
\end{itemize}
and for Tester, false negatives fell in four categories:
\begin{itemize}[leftmargin=*,nosep]
  \item \changes{\textbf{Misread semantics (19\%):} misunderstood IO relation.}
  \item \changes{\textbf{Output computation failure (58\%):} logic is described correctly but numerical results are wrong.}
  \item \changes{\textbf{Multiple valid answers (21\%):} tests assert one of several correct outputs (\Cref{tester_multiple_correct}).}
  \item \changes{\textbf{Insufficient coverage (2\%):} corner cases not tested.}
\end{itemize}

\changes{In \Cref{fig:tester_example}, the program (a greedy method with a suffix array) is correct. \textsc{Vanilla} misreads the role of \texttt{min(a[i],b[i])}; Tester predicts an incorrect output—consistent with findings that LLMs struggle to directly compute nontrivial results~\cite{gao2023pal}. \projName’s annotations make the purpose of \texttt{c} explicit (per-position minimum cost), enabling the correct verdict.}

\section{Threats of Validity}
\label{sec:threats}

LLMs are inherently nondeterministic~\cite{ouyang2025empirical}, leading to varied responses. To address this, we employed multiple reruns per sample with confidence-based sampling to ensure robust performance estimates. Although further reruns could slightly refine our results, the current methodology provides high-confidence evaluations.

Main threats to external validity are that the selected dataset may not generalize to other programs, or that performance enhancements will not generalize to other LLMs. To avoid data leakage, we constructed a new dataset from recent Codeforces competition problems, ensuring challenging, human-authored, and varied submissions. Despite efforts to minimize biases, future studies on alternative datasets could further validate generalizability and performance trends. Apart from that, real-world requirements contain ambiguities, which are typically not present in programming competition problems. Techniques like ClarifyGPT~\cite{mu2023clarifygpt} that detects ambiguities in informal requirements may address this concern.

\changes{We evaluated \projName across models of diverse sizes (7B to 72B) from 2 families (Qwen, Llama), including Qwen-32B-Coder, a model specialized for coding tasks.} This broad spectrum enhances generalizability; however, different LLM architectures may yield varying results. Nevertheless, observed trends suggest broad applicability, especially relevant given industry shifts toward proprietary small and mid-sized models~\cite{chen2024role}. Evaluating \projName on larger and more powerful models may help assess its scalability and effectiveness in leveraging more advanced reasoning capabilities.

\vspace{-2mm}

\section{Related Work}
\label{sec:related_work}

\projName introduces a new paradigm for reasoning about programs in NL. Its contributions are relevant to works on reasoning about code with LLMs, LLM-assisted testing and verification.


\textit{Reasoning About Code with LLMs.} Many generic approaches have been introduced to realize effective chain-of-though (CoT) reasoning~\cite{wei2022chain}. We employed Zero-shot-CoT~\cite{kojima2022large}, a popular automation of CoT, as a baseline in our experiments. CJ-Eval~\cite{zhao2024codejudge} uses LLMs as judges of program correctness applying an approach equivalent to our baseline Vanilla prompt (\Cref{sec:classification}). Since CJ-Eval is based on a widely-used Apps dataset~\cite{apps}, which is subject to data leakage, we constructed a new dataset \benchName. A large body of work focused on reasoning about concrete execution states. Zhang et al.~\cite{grounding} fine-tuned LLMs with concrete execution states to improve their ability to track execution faithfully. Similarly, NExT~\cite{ni2024next} enhances debugging using programs annotated with concrete execution traces. CodeMind~\cite{liu2024codemind} includes a ``specification reasoning'' task that measures how including test data in the specification helps the model generate correct code. REval~\cite{runtime} evaluates runtime behaviour reasoning capabilities, specifically, consistency in intermediate state prediction by analyzing variable states at specific execution points. The key difference of \projName is that it reasons about not individual executions but, potentially, infinitely many executions, without relying of specific inputs.

\textit{LLM-Assisted Formal Verification.} Recent works on LLM-assisted verification focused on translating natural language specifications into formal ones, and rely on external theorem provers or SMT solvers~\cite{endres2024can}. Marmaragan~\cite{spark} synthesizes formal Ada/SPARK contracts from natural language. Similarly, SynVer~\cite{symver} biases LLM-generated C code towards verifiability by producing assertions and invariants, subsequently verified by Coq or Verified Software Toolchain (VST). LLMLIFT~\cite{llmlift} integrates LLMs into verified compilation, generating proof sketches alongside code translation, with formal verification performed via SMT solvers. Stanford et al.~\cite{pc3} introduce Proof-Carrying Code Completions (PC3), prompting LLMs to produce Dafny~\cite{dafny} programs accompanied by machine-checkable correctness proofs, automating verification through Dafny. While these approaches embed formal verification into LLM-driven code generation pipelines, they fundamentally rely on external verification systems such as Dafny, Coq~\cite{Coq}, or Z3~\cite{z3}. Despite the promise of these methods, they have shown limited scalability, with Marmaragan achieving only 50.7\% correctness in its generated Ada/SPARK contracts \cite{spark} and Endres et al.~\cite{endres2024can} reporting that even with optimized prompting, only 77\% of LLM-generated postconditions were verifiable. \projName employs an alternative approach of constructing structured informal proofs in natural language, leveraging its expressive power and flexibility, for classifying program correctness. Following previous work~\cite{jiang2022draft} such informal proofs might be translated into formal ones to automate formal verification. Closest to our loop reasoning technique, Liu et al.~\cite{liu2024generalloopinvariantgeneration} propose LLM-SE, which combines symbolic execution with LLMs to generate loop invariants for memory-manipulating programs. While \projName does not target formal invariant synthesis, instead focusing in textual inductive reasoning. \changes{Although \projName does not provide formal guarantees, prior work shows that formal proofs mirror informal reasoning~\cite{jiang2022draft}, suggesting \projName could serve as a foundation for automated verification. Integrating \projName with techniques like symbolic CoT~\cite{xu2024faithful} may further enhance reasoning and provide guarantees. A promising future direction is to develop a hybrid (formal and natural) language to promote more grounding of the LLM's output}.

\textit{LLM-Based Automated Test Generation} Recent works~\cite{li2024large,schafer2023empirical,yuan2023no,chen2024chatunitest} have employed LLMs for automated test generation. Apart from that, the analysis of LLM-generated test outcomes has been used to facilitate program synthesis tasks~\cite{xiong2023program}, exemplified by CODET~\cite{chen2022codet}, AgentCoder~\cite{huang2023agentcoder}, CodeCoT~\cite{huang2023codecot}, and CoCoST~\cite{he2024cocost}. \Cref{sec:evaluation} shows that using LLM-generated tests for correctness classification, embodied in our \textit{Tester} baseline adopting AgentCoder's approach~\cite{huang2023agentcoder}, suffers from high false negative rate, as LLMs fail to infer precise assertions for complex requirements. Ideas of \projName could be synergized with testing to improve accuracy and cost. \newline
\hspace*{0.5em}\textit{Competition-Level Benchmarks} Recent studies reveal significant data leakage in widely used LLM benchmarks, inflating model performance due to test-training overlaps~\cite{training_on_bench,deng-etal-2024-unveiling}. New benchmarks, such as LiveCodeBench~\cite{jain2024livecodebenchholisticcontaminationfree}, CodeElo~\cite{quan2025codeelobenchmarkingcompetitionlevelcode}, ProBench~\cite{yang2025probenchbenchmarkinglargelanguage}, and USACO-based evaluations~\cite{shi2024language}, mitigate leakage by incorporating recent programming competition problems. However, they primarily assess code generation capabilities and do not provide solution submissions with correctness labels. In contrast, \benchName specifically targets correctness classification by offering labeled problem-solution pairs sourced from Codeforces 2024 contests, enabling evaluation of LLMs on their execution-state reasoning capabilities rather than memorization or test-based validation alone.

\vspace{-3mm}

\section{Conclusion}

This work introduces \projName, the first LLM-based program analysis approach to assess program correctness against natural language requirements. Unlike traditional verification and testing that depend on external tools or concrete test executions, \projName systematically applies natural language-based structural reasoning inspired by the strongest postcondition calculus and k-induction. Our evaluation shows significant performance improvements in classification quality, with \projName increasing the MCC by 61\% over Zero-shot-CoT, 72\% over Vanilla, and 106\% over an LLM testing-based classifier. Our novel k-induction further enhances performance, yielding an average MCC improvement of 26\% compared to a non-inductive approach. Although \projName consumes more tokens, the no-unroll variant provides a practical trade-off, achieving a 28\% MCC improvement over Zero-shot-CoT while using 90\% fewer tokens than the standard \projName version.

\begin{acks}
This work was sponsored by the National Key Research and Development Program of China under Grant No.\ 2022YFB4501902 and the National Natural Science Foundation of China (NSFC) under Grant No.\ W2411051.

\end{acks}

\bibliographystyle{ACM-Reference-Format}
\bibliography{references}

\onecolumn
\pagebreak
\twocolumn
\appendix

\section{Supplementary Materials}

All code, documentation, results and complimentary material for this work is available in our repository: \url{https://github.com/msv-lab/HoarePrompt}.\looseness=-1

\subsection{Details of Motivating Example}
\label{appendix:motivating_example}

For evaluating the motivating example presented in \Cref{sec:motivating_states}, we ran a series of experiments using three different \texttt{qwen2.5-instruct} models (\texttt{7b}, \texttt{coder-32b}, \texttt{72b}) \changes{and one Llama model \\(\texttt{Llama3.1-70b-instruct})} at three temperatures \((0.5, 1.0, 1.5)\).\\ For each temperature, we tested:
\begin{itemize}
    \item \textbf{Not Annotated}: 6 different prompts with no reachable state annotations.
    \item \textbf{Annotated}: 6 different prompts that included reachable state descriptions.
\end{itemize}
Each model/prompt combination was run 10 times, for a total of 1080 calls across all setups. Table~\ref{appendix:tab:motivating_example_results} shows the success rates (the percentage of calls that correctly identified the code as buggy).

\begin{table}[ht]
\centering
\caption{Success rates (\%) for detecting the indentation bug under different prompts (``Not Annotated'' vs.~``Annotated''), temperatures, and models. Each cell shows \emph{successful runs} / \emph{total runs} (percentage).}
\label{appendix:tab:motivating_example_results}
\begin{tabular}{l c c}
\toprule
\multicolumn{3}{c}{\textbf{Temperature = 0.5}} \\
\midrule
\textbf{Model} & \textbf{Not Annotated} & \textbf{Annotated} \\
\midrule
\texttt{7b-instruct}          & 0/60 (0\%)   & 7/60 (11.7\%) \\
\texttt{coder-32b-instruct}   & 9/60 (15\%)  & 51/60 (85\%)  \\
\texttt{72b-instruct}         & 21/60 (35\%) & 51/60 (85\%)  \\
\changes{\texttt{llama3.1-70b-instruct}}  & \changes{30/60 (50\%)} & \changes{60/60 (100\%)}  \\
\midrule
\multicolumn{3}{c}{\textbf{Temperature = 1.0}} \\
\midrule
\textbf{Model} & \textbf{Not Annotated} & \textbf{Annotated} \\
\midrule
\texttt{7b-instruct}          & 0/60 (0\%)   & 2/60 (3.3\%)  \\
\texttt{coder-32b-instruct}   & 9/60 (15\%)  & 50/60 (83.3\%)\\
\texttt{72b-instruct}         & 29/60 (48.3\%) & 49/60 (81.7\%)\\
\changes{\texttt{llama3.1-70b-instruct}}  & \changes{30/60 (50\%)} & \changes{60/60 (100\%)}  \\

\midrule
\multicolumn{3}{c}{\textbf{Temperature = 1.5}} \\
\midrule
\textbf{Model} & \textbf{Not Annotated} & \textbf{Annotated} \\
\midrule
\texttt{7b-instruct}          & 4/60 (6.7\%)  & 5/60 (8.3\%)  \\
\texttt{coder-32b-instruct}   & 7/60 (11.7\%) & 51/60 (85\%)  \\
\texttt{72b-instruct}         & 31/60 (51.7\%) & 42/60 (70\%)  \\
\changes{\texttt{llama3.1-70b-instruct}}  & \changes{40/60 (67\%)} & \changes{60/60 (100\%)}  \\
\bottomrule
\end{tabular}
\end{table}

\subsection{Natural Language and LLMs for State Tracking}
\label{appendix:NL}

We explore the effectiveness of Qwen2.5-72B-Instruct in inferring the final state (postcondition) of a program segment by leveraging semantic context and real-world knowledge. Unlike formal verification methods that rely on explicit rules and symbolic execution, Qwen2.5-72B-Instruct can generate intuitive and human-readable explanations of code behavior. This approach enhances analysis precision by utilizing variable names and structure to improve reasoning. While it does not provide formal correctness guarantees, it acts as a practical fuzzy correctness classifier, making it useful for tasks like bug detection and code generation.

\begin{figure}[t]
    \centering
    \includegraphics[width=0.8\linewidth]{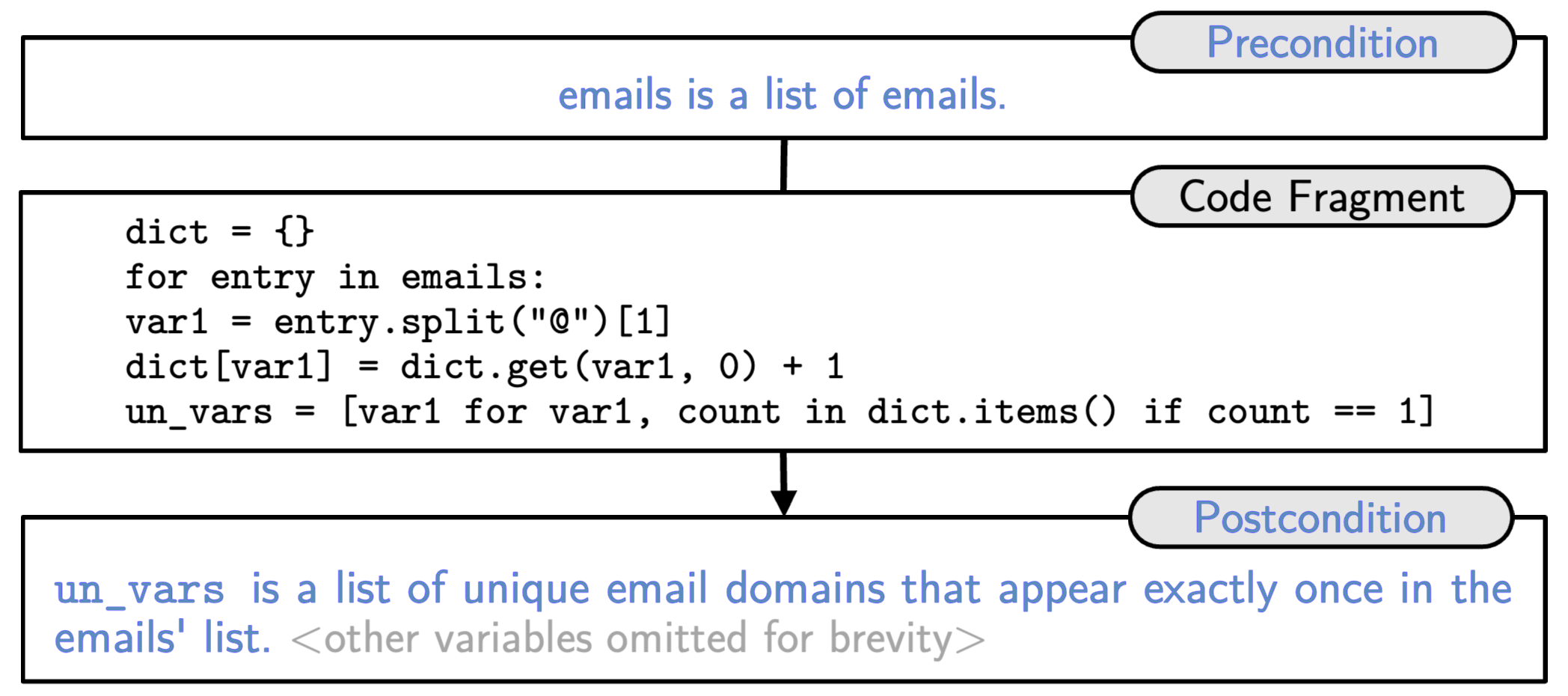}
    \caption{Example 2 - Unique Email Domains \label{fig:unique_email}}
    \vspace{-2mm}
\end{figure}

\begin{figure}[t]
    \centering
    \includegraphics[width=0.8\linewidth]{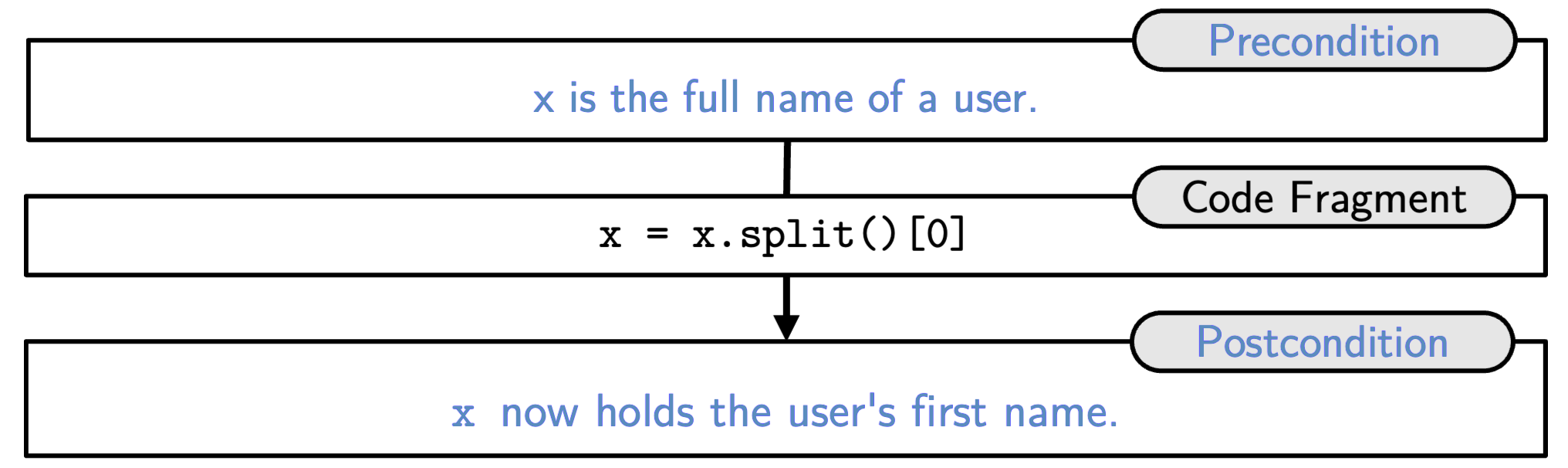}
    \caption{Example 3 - First name extraction \label{fig:first_name}}
    \vspace{-2mm}
\end{figure}

In Figures \ref{fig:unique_email},\ref{fig:first_name} we present two more examples to complement that in \Cref{sec:motivating_NL}. These examples illustrate how an LLM can concisely infer express program states by leveraging natural language. We believe this ability of LLM to succinctly capture program behavior using user-centric terms helps \projName to accurately classify program correctness by bridging the gap between program semantics and informal requirements.

\subsection{Dataset Details}

\begin{figure}
    \centering
    \includegraphics[width=0.9\columnwidth]{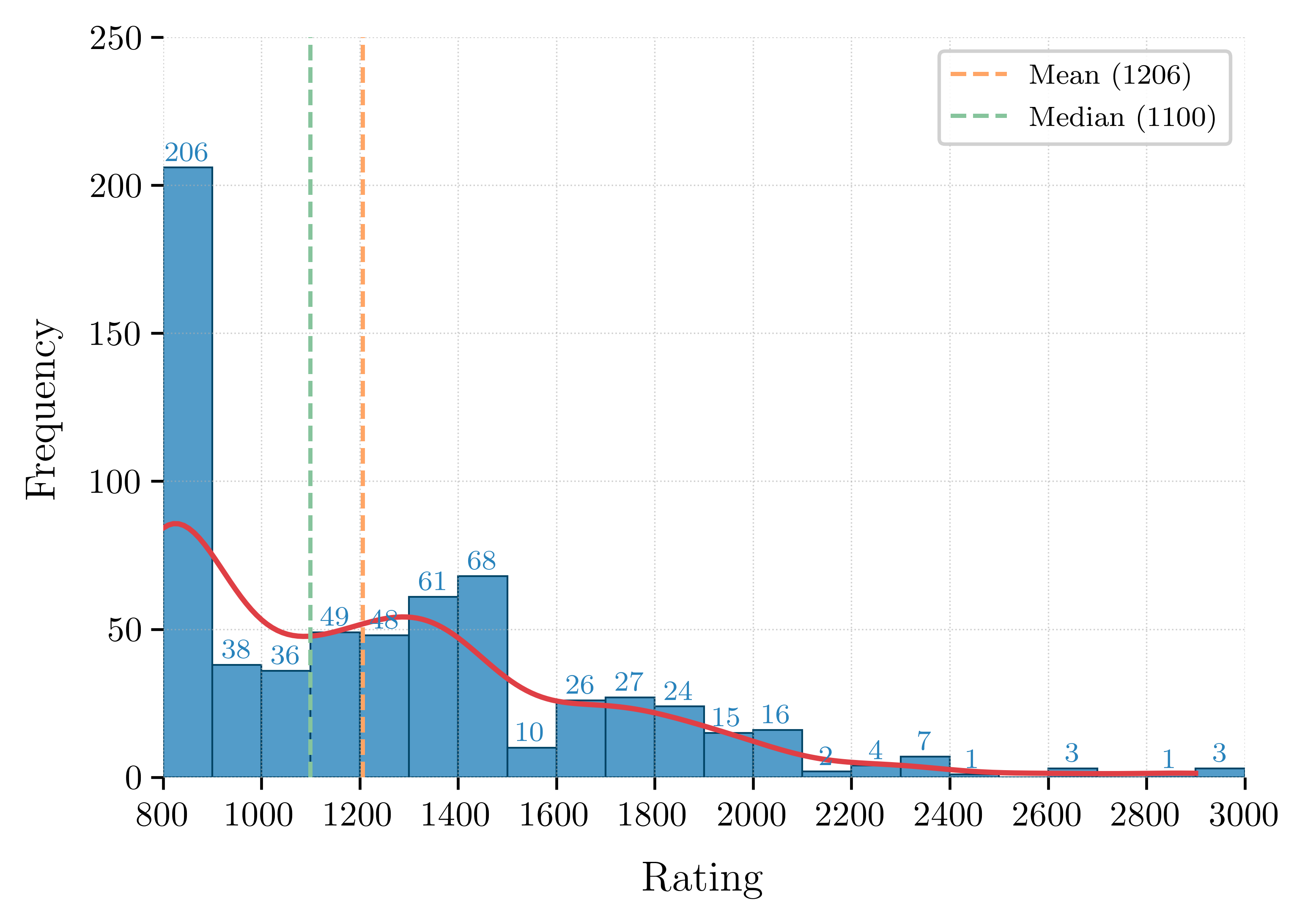}  
    \caption{The Frequency Histogram of Rating for \benchName with x-axis representing the rating intervals of the problems and y-axis representing the number of problem-code pairs within each rating interval in the dataset.}
    \label{fig:rating_distribution}
    \vspace{-3mm}
\end{figure}

\label{dataset_details}

To better illustrate the difficulty of problems in \benchName, we use the \emph{rating} tags for problems provided by Codeforces as an indicator. Initially, rating was solely an indicator of user rankings on programming competition platforms. Subsequently, to better assist users in selecting appropriate problems for practice, Codeforces began assigning each problem a rating to represent its difficulty, which is determined by the solving statistics of participants for this problem~\cite{codeforces2019problem}. Formally, a problem's rating $x$ quantifies that participants with a rating of $x$ have a 50\% probability of solving it on their first encounter.

Rating for problems ranges from 800 to 3500 and serves as a difficulty coefficient indicator, where a higher rating indicates greater problem difficulty. During the problem collection process, we also gathered the ratings of each problem and present the rating distribution of our dataset. As shown in Figure~\ref{fig:rating_distribution}, our dataset has an average rating of 1206. According to the statistics and evaluations from CodeElo\cite{quan2025codeelobenchmarkingcompetitionlevelcode}, a problem with a rating of 1206 indicates that approximately 60\% of all participants on Codeforces find it challenging (with less than a 50\% probability of solving it on their first encounter), and only 2 of the 33 tested LLMs achieved a higher rating than 1206, which is difficult for LLMs to simply retrieve.

Additionally, as can be seen from the figure, the data presents a skewed distribution of "higher on the left and lower on the right". The frequency in low rating intervals (such as 800–1000) is significantly higher, and as the rating increases, the frequency generally shows a downward trend. The main reason is that as the difficulty of the problem increases (the rise in rating), the number of people who solve this problem (or even the number of people who attempt to solve it) decreases. Therefore, there are fewer submission records available for sampling, and even fewer submission records that meet our sampling strategy.

The scraped problems were originally in raw HTML format. We parsed out all the useful content w.r.t. the problem and converted it into Markdown format (to be consistent with our prompt), which may include the problem description, input format, output format, examples, notes, and so on, to ensure the sufficiency of the problem information and details.

For a problem, we filtered out all its Python submissions. The specific strategy for selecting submissions is as follows:
\begin{itemize}
    \item Include three consecutive incorrect/correct code pairs (each pair comes from the same user). If there are less than three, include all that are available. If the problem has not been solved by anyone, skip it. 
    \item If available, include one code with a test passed rate of 50\% (or close to 50\% due to an odd number of test cases), and one code with only a single failing test case. 
\end{itemize}

\changes{Furthermore, we provide a detailed comparison in Table~\ref{tab:dataset_comparison} of our \benchName~dataset to two prominent datasets: LiveCodeBench~\cite{jain2024livecodebenchholisticcontaminationfree} and APPS~\cite{apps}, based on key metrics such as solution complexity, code length, loop depth, and description verbosity.}

\setlength{\tabcolsep}{4pt}
\begin{table}[h]
\centering
\begin{tabular}{|l|c|c|c|c|c|c|}
\hline
\changes{\textbf{Dataset}} & \changes{\textbf{LOC}} & \changes{\textbf{LD}} & \changes{\textbf{DL}} & \changes{\textbf{CC}} & \changes{\textbf{HD}} & \changes{\textbf{HV}} \\
\hline
\changes{\textbf{\benchName}}      & \changes{25.1}  & \changes{1.29} & \changes{2156.83} & \changes{7.77}  & \changes{28.8}  & \changes{1063.2} \\
\changes{\textbf{LiveCodeBench}}   & \changes{9.95}  & \changes{1.01} & \changes{42.35}   & \changes{4.18}  & \changes{21.81} & \changes{416.2}  \\
\changes{\textbf{APPS}}            & \changes{24.03} & \changes{1.29} & \changes{1416.06} & \changes{7.63}  & \changes{34.12} & \changes{1103.6} \\
\hline
\end{tabular}
\caption{\changes{Comparison of dataset complexity based on Lines of Code (LOC), Loop Depth (LD), Description Length (DL), Cyclomatic Complexity (CC), Halstead Difficulty (HD), and Halstead Volume (HV). Higher CC, HD, and HV values indicate more complex programs.}}
\label{tab:dataset_comparison}
\end{table}
\setlength{\tabcolsep}{6pt}

\subsection{Rerun Strategy and Variance Control}
\label{appendix:reruns}

\changes{To ensure statistical reliability given LLM response variability, we computed the number of reruns required to achieve a high-confidence estimate with low error. Our approach follows the framework of Ouyang et al.~\cite{ouyang2025empirical} and the confidence-weighted aggregation method of Xiong et al.~\cite{xiong2024llmsexpressuncertaintyempirical}.}

\paragraph{\changes{Confidence Estimation.}}  
\changes{Given a binary prediction (Correct or Incorrect), the LLM assigns a log-probability \( \log P(Y) \), which is converted to confidence via \( \exp(\log P(Y)) \). We aggregate confidences across \( M \) reruns using the Avg-Conf method:}
\[
\changes{
C_{\text{avg-conf}} = \frac{\sum_{i=1}^{M} \mathbb{I}(Y_i = Y) \cdot C_i}{\sum_{i=1}^{M} C_i}
}
\]
\changes{where \( Y_i \) is the $i$-th response, \( C_i \) its confidence score, and \( \mathbb{I}(Y_i = Y) \) is an indicator for agreement with the majority response.}

\paragraph{\changes{Rerun Count Calculation.}}  
\changes{Let \( P \) be the estimated consistency of the LLM across reruns, \( X \) the number of evaluation instances, and \( R \) the number of reruns. We aim to bound the margin of error \( \epsilon \) for a desired confidence level \( C \), modeled with a binomial distribution:}
\[
\changes{
\epsilon = Z \sqrt{\frac{P(1-P)}{N}}, \quad N = R \cdot X, \quad R = \frac{Z^2 P(1-P)}{X \cdot \epsilon^2}
}
\]
\changes{where \( Z \) is the critical value for the desired confidence level (e.g., \( Z = 2.33 \) for 98\%).}

\paragraph{\changes{Final Parameters.}}  
\changes{
Using:
\begin{itemize}
    \item Dataset size: \( X = 645 \),
    \item Observed average confidence: \( C = 0.963 \),
    \item Desired confidence level: 98\% (\( Z = 2.33 \)),
    \item Error margin: \( \epsilon = 0.01 \),
\end{itemize}
we compute \( R \approx 3 \), meaning all experiments are repeated three times to achieve statistically reliable results.
}

\subsection{Example: Tester Misclassification due to multiple correct solutions}
\label{tester_multiple_correct}

\changes{This example from our dataset \benchName{}, as seen on Figure \ref{multiple_correct}, highlights a common limitation of the LLM-generated tester. Specifically, it shows how problems with multiple valid outputs can lead the tester to incorrectly reject a correct implementation.}
\begin{figure}[h]

\caption{\changes{Example of problem with multiple correct solutions}}
    \centering
    \label{multiple_correct}
    \begin{minipage}[t]{0.31\textwidth}
        \vspace{0pt}
        \begin{tcolorbox}[colframe=darkgreen!100, colback=gray!3, title=Problem Description, fonttitle=\centering, boxrule=0.7mm, sharp corners=southwest, boxsep=1pt]
        \footnotesize
Given integer $x$, find any $y$ ($1 \le y < x$) such that
\[
\gcd(x,y) + y
\]
is maximized.

Multiple $y$ may be equally valid.

        \end{tcolorbox}
    \end{minipage}
    \hfill
    \begin{minipage}[t]{0.31\textwidth}
        \vspace{0pt}
        \begin{tcolorbox}[colframe= softblue, colback=gray!3, title=Tester Generated Test Case, fonttitle=\centering, boxrule=0.7mm, sharp corners=southwest, boxsep=1pt]
        \footnotesize

\textbf{Input:}
\begin{verbatim}
1
10
\end{verbatim}
\textbf{Output:}
\begin{verbatim}
5
\end{verbatim}

\[
\gcd(10,5)=5
\quad 5+5=10
\]

Tester selected $y=5$ as the solution.
        \end{tcolorbox}
    \end{minipage}
    \hfill
    \begin{minipage}[t]{0.31\textwidth}
        \vspace{0pt}
        \begin{tcolorbox}[colframe=nicered!90, colback=gray!3, title=Tester Classification, fonttitle=\centering, boxrule=0.7mm, sharp corners=southwest, boxsep=1pt]
        \footnotesize
        \textbf{Incorrect}\\
Our implementation produced:

\begin{verbatim}
9
\end{verbatim}

\[
\gcd(10,9)=1,\;1+9=10
\]

Also optimal, but tester flagged as \textbf{incorrect} due to strict expected match.
        \end{tcolorbox}
    \end{minipage}
\end{figure}

\subsection{Example : Impact of Incorrect Test Cases on LLM-Driven Code Refinement based on Agent Coder}
\label{ex_agent_coder_test}

\changes{In our experiments with Llama3.1, we observed a striking example of how incorrect test cases can mislead LLM-based program synthesis or correction systems, resulting in previously correct implementations being erroneously altered.}

\paragraph{\changes{Context.}} \changes{The underlying programming problem involves a little boy, Nikita, who builds a tower using cubes. The process follows two simple rules:}
\begin{itemize}
    \item \changes{In each of $n$ moves, Nikita must either \emph{add exactly one cube} on top of the tower or \emph{remove exactly one cube} from the top.}
    \item \changes{The tower starts with $0$ cubes, and it is never allowed to have a negative height—removing a cube from an empty tower is invalid.}
\end{itemize}

\changes{Given this setup, the task is to determine whether it is possible, after exactly $n$ moves, to end up with a tower that has exactly $m$ cubes.}

\changes{Two simple conditions fully characterize the problem:}
\begin{itemize}
    \item \changes{$m \le n$, since each move changes the height by at most $1$.}
    \item \changes{$(n - m)$ is even, ensuring moves can be paired into balanced ``add and remove'' operations to reach exactly $m$.}
\end{itemize}
\changes{Thus, the tower ends with $m$ cubes after $n$ moves if and only if:}
\[
\changes{
\text{possible} \Longleftrightarrow \left(m \le n \;\;\wedge\;\; (n - m) \bmod 2 = 0\right)
}
\]
\changes{The LLM produced an implementation which faithfully reflects this reasoning:}
\begin{verbatim}
def solve(n, m):
    if m > n:
        return "No"
    if (n - m) % 2 == 0:
        return "Yes"
    else:
        return "No"
\end{verbatim}

\changes{\paragraph{Faulty Test Cases.} LLaMA3.1-70B generated test cases that the expected output was incorrectly calculated by the model and they were subsequently used as feedback to refine the code.Two such test cases are:}
\begin{quote}
\begin{itemize}
    \item \changes{\texttt{Input: 1 0}, \texttt{Expected Output: Yes}}
    \item \changes{\texttt{Input: 5 0}, \texttt{Expected Output: Yes}}
\end{itemize}
\end{quote}
\changes{which contradict the problem’s constraints: for example, with \texttt{1 0}, it is impossible to end with 0 cubes after a single move since removing a cube is invalid on an empty tower.}
\paragraph{\changes{Misguided Corrections.}} 
\changes{The faulty test cases were executed on the original correct implementation, resulting in output mismatches. Interpreting these mismatches as genuine errors, the LLM attempted to fix the code to satisfy the flawed expectations—introducing ad-hoc conditions that deviated from the correct logic:}

\begin{verbatim}
if (n - m) % 2 == 0 or ((n - m) % 2 == 1 and m != 0):
    return "Yes"
\end{verbatim}
\changes{which breaks the parity logic and admits invalid solutions, thereby turning correct code into incorrect code.}

\subsection{Example: HoarePrompt-Guided Correction for Code Generation based on Agent Coder}
\label{ex_hoareprompt_beautiful_pairs}

\paragraph{\changes{Context.}}  
\changes{Count all pairs \(\langle i,j\rangle\) in an array \(a\) so that}
\[
\changes{
a_i + a_j \equiv 0 \pmod{x}
\quad\text{and}\quad
a_i - a_j \equiv 0 \pmod{y}.
}
\]

\paragraph{\changes{Faulty Implementation.}}
\begin{verbatim}
for num in a:
    rx = num % x
    ry = num % y
    remainder = (rx + ry) % y
    complement = ((x - rx) % x + ry) % y
    beautiful_pairs += cnt[complement]
    cnt[remainder] += 1
\end{verbatim}

\paragraph{\changes{HoarePrompt Annotation.}}  
\changes{HoarePromptwith Llama3.1,highlights the line}
\begin{verbatim}
    remainder = (rx + ry) % y
\end{verbatim}
\changes{with the note}  
\changes{\emph{the remainder is the remainder of the sum of num divided by x plus the remainder of num divided by y, divided by y}}

\paragraph{\changes{Feedback.}}  
\changes{“The program contains a bug. This bug stems from merging \(\texttt{num}\bmod x\) and \(\texttt{num}\bmod y\) into one \(\bmod y\) expression, dropping the independent \(\bmod x\) check.”}

\paragraph{\changes{Corrected Implementation.}}
\begin{verbatim}
cnt = defaultdict(int)
for num in a:
    r_x = num % x
    r_y = num % y
    need_x = (-r_x) % x
    beautiful_pairs += cnt[(need_x, r_y)]
    cnt[(r_x, r_y)] += 1
\end{verbatim}
\changes{Here each element is bucketed by the pair \((r_x,r_y)\), so “sum $\equiv$ 0 mod x” and “difference $\equiv$ 0 mod y” are enforced independently.}

\subsection{MCC Calculation and Justification}
\label{appendix:mcc}

To evaluate the quality of binary correctness classification, we adopt the \textbf{Matthews Correlation Coefficient (MCC)} as our primary metric. Given the confusion matrix entries—true positives (TP), true negatives (TN), false positives (FP), and false negatives (FN)—the MCC is computed as:

\[
\text{MCC} = \frac{(\text{TP} \times \text{TN}) - (\text{FP} \times \text{FN})}{\sqrt{(\text{TP} + \text{FP})(\text{TP} + \text{FN})(\text{TN} + \text{FP})(\text{TN} + \text{FN})}}
\]

The MCC yields a score in \([-1, 1]\), where:
\begin{itemize}
    \item \(1\) indicates perfect prediction,
    \item \(0\) indicates random prediction,
    \item \(-1\) indicates total disagreement.
\end{itemize}

We chose MCC over commonly used alternatives such as ROC AUC or accuracy because MCC takes into account \emph{all four} confusion matrix categories, and it remains informative even under class imbalance. This makes it particularly suitable for correctness classification, where false negatives (missing bugs) and false positives (flagging correct code as incorrect) are both costly.  As shown by Chicco and Jurman~\cite{mcc}, ROC AUC can yield misleadingly high scores even for classifiers with low precision or poor practical performance. Moreover, ROC AUC considers performance over all thresholds, not the concrete operating point used for binary decisions. In contrast, MCC directly reflects classifier behavior at the applied threshold and ensures high values only when \emph{sensitivity, specificity, precision, and NPV} are all high.
Therefore, we follow their recommendation that MCC should replace ROC AUC as the standard metric in binary classification settings~\cite{mcc}, especially in sensitive domains like program verification, where balanced correctness evaluation is critical.

\onecolumn  

\section{LLM Prompts}

In this section of the Appendix, we include all prompts provided to the LLMs during the evaluation and experiments in our study. Where FSL was employed, the FSL examples are also included.

\subsection{Zero-Shot Chain-of-Thought (CoT) Prompt}
\label{0-shot-COT-prompt}
\begin{tcolorbox}[ colback=gray!10, colframe=black, arc=5pt]
\begin{lstlisting}[language=, basicstyle=\ttfamily\footnotesize]
Your task is to determine if a given Python program is correct based on the provided problem description. Assume valid inputs as described in the problem description.

First, explain your reasoning step by step, then reply with:  
- Correctness: `True` if the given program is correct.  
- Correctness: `False` if the given program is incorrect.


Problem:  
{decription}

Program:  
{code}

# Your response:  
Reasoning:
Correctness: **True** or **False**

\end{lstlisting}
\end{tcolorbox}

\subsection{Vanilla Prompt (No Chain-of-Thought)}
\label{vanilla-prompt}
\begin{tcolorbox}[ colback=gray!10, colframe=black, arc=5pt]
\begin{lstlisting}[language=, basicstyle=\ttfamily\footnotesize]
Your task is to determine if a given Python program is correct based on the provided problem description. Assume valid inputs as described in the problem description.

Reply with:  
- Correctness: `True` if the given program is correct.  
- Correctness: `False` if the given program is incorrect.


Problem:  
{description}

Program:  
{code}

# Your response:  
Correctness: **True** or **False**

\end{lstlisting}
\end{tcolorbox}

\subsection{Prompt for Input-Output Test Case Generation}
\begin{tcolorbox}[ colback=gray!10, colframe=black, arc=5pt]
\begin{lstlisting}[language=, basicstyle=\ttfamily\footnotesize]
**Role**: As a tester, your task is to create comprehensive test cases for the following coding problem. These test cases should encompass Basic and Edge scenarios to ensure the code's robustness, reliability, and scalability.

**Problem Description**:  
{description}

**1. Basic Test Cases**:  
- **Objective**: To verify the fundamental functionality of the `has_close_elements` function under normal conditions.

**2. Edge Test Cases**:  
- **Objective**: To evaluate the function's behavior under extreme or unusual conditions.

**Instructions**:  
- Implement a comprehensive set of test cases following the guidelines above.  
- Ensure each test case is complete (no omission) and well-documented with comments explaining the scenario it covers.  
- Pay special attention to edge cases as they often reveal hidden bugs.  
- Do not repeat, do not summarize.  

All test cases you give need to strictly follow the problem description and format like this:

# Test 1  
**Input**:  
```

```
**Output**:  
```

```

\end{lstlisting}
\end{tcolorbox}

\subsection{HoarePrompt Correctness Classification Prompt}
\label{correctness-single-func-prompt}
\begin{tcolorbox}[ colback=gray!10, colframe=black, arc=5pt]
\begin{lstlisting}[language=, basicstyle=\ttfamily\footnotesize]
Your task is to determine if a given Python program is correct based on the problem description and the execution states of the program provided as comments. Assume valid inputs as described in the problem description.

First, explain your reasoning, then reply with:  
- Correctness: `True` if the given program is correct.  
- Correctness: `False` if the given program is incorrect.


**# Problem:**  
{description}

**# Annotated Program:**  
{annotated_program}

**# Your response:**  
Reasoning:  
Correctness: **True** or **False**

\end{lstlisting}
\end{tcolorbox}

\subsection{HoarePrompt Correctness Classification for Multi-Function Programs}
\label{correctness-multi-func-prompt}
\begin{tcolorbox}[ colback=gray!10, colframe=black, arc=5pt]
\begin{lstlisting}[language=, basicstyle=\ttfamily\footnotesize]
Your task is to determine if a given Python program is correct based on the problem description and the execution states of the program provided as comments. Assume valid inputs as described in the problem. The program is made of multiple functions and the program is **correct** only if all its functions together meet the problem description.

First, explain your reasoning, then reply with:  
- Correctness: `True` if the given program is correct.  
- Correctness: `False` if the given program is incorrect.


**# Problem:**  
{description}

**# Annotated Functions:**  
{functions}

**# Your response:**  
Reasoning:  
Correctness: **True** or **False**

\end{lstlisting}
\end{tcolorbox}
\flushbottom

\subsection{Precondition Extraction Prompt}
\label{pre_prompt}
\begin{tcolorbox}[ colback=gray!10, colframe=black, arc=5pt]
\begin{lstlisting}[language=, basicstyle=\ttfamily\footnotesize]
You are given a programming problem description and a function definition for a function that solves this problem. From the problem description, extract a description of the values of the program's input variables and the relationships between these variables. We refer to this description as the **precondition**. Print the precondition following the word **"Precondition"**, and surround it with double asterisks (**). Follow these examples:

### Example 1  
**Problem description:** Write a function to find the minimum cost path to reach (m, n) from (0, 0) for the given cost matrix cost[][] and a position (m, n) in cost[][].  
**Function definition:**
```python
def min_cost(cost, m, n):
```
**Precondition:** **cost is a 2D list of non-negative integers, m and n are non-negative integers such that 0 <= m < len(cost) and 0 <= n < len(cost[0]).**

### Example 2  
**Problem description:** Write a function to find the similar elements from the given two tuple lists.  
**Function definition:**
```python
def similar_elements(test_tup1, test_tup2):
```
**Precondition:** **test_tup1 and test_tup2 are tuples.**

### Example 3  
**Problem description:** Write a Python function to identify non-prime numbers.  
**Function definition:**
```python
def is_not_prime(n):
```
**Precondition:** **n is an integer greater than 1.**

### Example 4  
**Problem description:** Write a function to find the largest integers from a given list of numbers using the heap queue algorithm.  
**Function definition:**
```python
def heap_queue_largest(nums, n):
```
**Precondition:** **nums is a list of integers, and n is a non-negative integer such that 0 <= n <= len(nums).**

### Example 5  
**Problem description:** Write a function to find the number of ways to fill it with 2 x 1 dominoes for the given 3 x n board.  
**Function definition:**
```python
def count_ways(n):
```
**Precondition:** **n is a non-negative integer.**

### Example 6  
**Problem description:** Find the average of the positive integers in a list.  
**Function definition:**
```python
def func_1(numbers):
```
**Precondition:** **numbers is a list of integers.**

### **Your Task**

**Problem description:**  
{description}

**Function definition:**
```python
{program}
```

\end{lstlisting}
\end{tcolorbox}
\clearpage
\subsection{Precondition Extraction Prompt for Multi-Function Programs}
\label{pre_multi_prompt}
\begin{tcolorbox}[ colback=gray!10, colframe=black, arc=5pt]
\begin{lstlisting}[language=, basicstyle=\ttfamily\footnotesize]
You are given a programming problem description and a function that contributes to the solution of this problem. The total solution comprises multiple functions, and this is just one of them. 
From the problem description, and based on the variables used in the signature of this specific function, extract a description of the values of the variables in the function signature and the relationship between them. We refer to this description as the **precondition**. Print the precondition following the word **Precondition**, and surround it with double asterisks (**). Follow these examples:
Remember, the function given may not solve the problem directly but may perform a side functionality that contributes to the total solution. Include information only about the variables in the function signature.

---

### Example 1  
**Problem description:** Write a function to find the minimum cost path to reach (m, n) from (0, 0) for the given cost matrix cost[][] and a position (m, n) in cost[][].  
**Program:**
```python
def min_cost(cost, m, n):
    tc = [[0 for x in range(C)] for x in range(R)]
    tc[0][0] = cost[0][0]
    for i in range(1, m+1):
        tc[i][0] = tc[i-1][0] + cost[i][0]
    for j in range(1, n+1):
        tc[0][j] = tc[0][j-1] + cost[0][j]
    for i in range(1, m+1):
        for j in range(1, n+1):
            tc[i][j] = min(tc[i-1][j-1], tc[i-1][j], tc[i][j-1]) + cost[i][j]
    return tc[m][n]
```
**Precondition:** **cost is a 2D list of non-negative integers, m and n are non-negative integers such that 0 <= m < len(cost) and 0 <= n < len(cost[0]).**

---

### Example 2  
**Problem description:** Write a function to find the similar elements from the given two tuple lists.  
**Program:**
```python
def are_similar(elem, elem1):
    if elem == elem1:
        return True
    else:
        return False
```
**Precondition:** **elem1 and elem are values of any type and value.**

---

### Example 3  
**Problem description:** Write a Python function to identify if 2 consecutive integers in a list are not prime.  
**Program:**
```python
import math
def is_not_prime(n):
    result = False
    for i in range(2, int(math.sqrt(n)) + 1):
        if n % i == 0:
            result = True
    return result
```
**Precondition:** **n is an integer greater than 1.**

\end{lstlisting}
\end{tcolorbox}

\begin{tcolorbox}[ colback=gray!10, colframe=black, arc=5pt]
\begin{lstlisting}[language=, basicstyle=\ttfamily\footnotesize]
### Example 4 
**Problem description:** Write a function to find the largest integers from a given list of numbers using the heap queue algorithm.  
**Program:**
```python
import heapq as hq
def heap_queue_largest(nums, n):
    largest_nums = hq.nlargest(n, nums)
    return largest_nums
```
**Precondition:** **nums is a list of integers, and n is a non-negative integer such that 0 <= n <= len(nums).**

---

### **Your Task**

**Problem description:**  
{description}

**Program:**
```python
{program}
```

\end{lstlisting}
\end{tcolorbox}
\clearpage
\flushbottom
\subsection{Simple Statement Execution Prompt}
\label{simple_statement_prompt}
\begin{tcolorbox}[ colback=gray!10, colframe=black, arc=5pt]
\begin{lstlisting}[language=, basicstyle=\ttfamily\footnotesize]
You have been assigned the role of a program executor, responsible for simulating the execution of Python code. You will be provided with an initial state and a Python code snippet. You need to provide the output state after running the Python code based on the initial state. Please avoid describing how the program runs. When a variable has a specific value, use that specific value directly for calculations. If a return takes place, make sure to always mention that a value or variable has been returned in the output state. You must adhere to the text format: **Output State: `output state`**.

Include all the information of the precondition that is still valid after the code has been executed. Just update the values of the variables that have been changed by the code.

I am giving you some examples to understand the task better. Then I am giving you your task.

---

### Example 1  
**Initial State:** `str` is a string  
```python
n = int(input())
```
**Output State:** **`str` is a string, `n` is an input integer**

---

### Example 2  
**Initial State:** Variables can hold any values  
```python
i += 1
```
**Output State:** **variable `i` is increased by 1**

---

### Example 3  
**Initial State:** `n` is either 3 or 5  
```python
m = n + 1
```
**Output State:** **`n` is either 3 or 5; `m` is either 4 or 6**

---

### Example 4  
**Initial State:** `x` is a positive integer; if `x` is greater than 10, then `z = 0`, else `z = 1`.  
```python
y = -z
```
**Output State:** **`x` is a positive integer, if `x` is greater than 10 then `z=0` and `y=0`, else `z=1` and `y=-1`**

---

### Example 5  
**Initial State:** `total` is 0, `i` is 1  
```python
break
```
**Output State:** **`total` is 0, `i` is 1 and we break out of the most internal loop or if statement.**

\end{lstlisting}
\end{tcolorbox}

\begin{tcolorbox}[ colback=gray!10, colframe=black, arc=5pt]
\begin{lstlisting}[language=, basicstyle=\ttfamily\footnotesize]
### Example 6  
**Initial State:** `total` is positive, `num` is negative, `x` is 0  
```python
x = total - num
```
**Output State:** **`total` is positive, `num` is negative, `x` is a positive value equal to `total - num`.**

### **Your Task**

**Initial State:**  
{pre}

```python
{program}
```

\end{lstlisting}
\end{tcolorbox}

\subsection{Compound Simple Statements Execution Prompt}
\label{compund statement prompt}
\begin{tcolorbox}[ colback=gray!10, colframe=black, arc=5pt]
\begin{lstlisting}[language=, basicstyle=\ttfamily\footnotesize]
You have been assigned the role of a program executor, responsible for simulating the execution of Python code. You will be provided with an initial state and a Python code snippet consisting of multiple lines. Your task is to execute all the lines in sequence and provide the output state after the entire code block has been run. Avoid describing how the program runs step-by-step for individual lines but instead focus on the combined effect of all lines. When a variable has a specific value, use that specific value directly for calculations.

Include all the information from the precondition that remains valid after the code execution and update the values of any variables that are modified by the code. Provide the final state, including the state of all the variables after the execution of the code snippet. Use the text format: **Output State: `output state`**.

Here are some examples to help you understand the task:

---

### Example 1  
**Initial State:** `a` is 5, `b` is 3  
```python
c = a + b
d = c * 2
```
**Output State:** **a is 5, b is 3, c is 8, d is 16**

---

### Example 2  
**Initial State:** `x` is a positive integer  
```python
n = int(input())
x += n
```
**Output State:** **x is a positive integer equal to its original value plus `n`, `n` is an integer**

---

### Example 3  
**Initial State:** `n` is 3, `total` is 1  
```python
total += n
pass
n = max(total, n)
```
**Output State:** **n is 4, total is 4**

---

### **Your Task**

**Initial State:**  
{pre}

```python
{program}
```

Now, please analyze the entire block of code and provide the final output state. List the impact of all lines on the program, check the previous values of affected variables, and calculate the states of the variables after the code executes. Be as specific as possible, combining changes from all lines into a single coherent final state. Include all valid information from the precondition and update only what is modified by the code.

In your response, strictly use the format: **Output State: `the output state you calculate`.**, and describe this output state in natural language easily understandable by humans.

\end{lstlisting}
\end{tcolorbox}

\subsection{Print Commands Execution Prompt}
\label{print_prompt}
\begin{tcolorbox}[ colback=gray!10, colframe=black, arc=5pt]
\begin{lstlisting}[language=, basicstyle=\ttfamily\footnotesize]
You will be given an **initial state** (precondition) and a **Python code snippet** containing a `print` statement. Your task is to **determine exactly what will be printed** when the statement executes.

- If a variable or object has a known **explicit value**, use that value in the output.  
- If a variable or object is defined by a **formula or condition**, describe its value using the given information.  
- Always provide the most **precise** description possible based on the precondition.
- Format the final output as: **Output: `what is printed`**.

I am giving you some examples to understand the task better. Then I am giving you your task.

---

### Example 1  
**Initial State:** `arr` is a list containing 1, 2, 3, 4, 5, and `sum` is the sum of all elements in the list `arr`.  
```python
print(arr[2], sum)
```
**Output:** **3, sum (where sum is the sum of all elements in list `arr`)**

---

### Example 2  
**Initial State:** The `points` list is a list of points. The `shoelace_sum` is the sum of all terms calculated as \(x_1 * y_2 - y_1 * x_2\) for each consecutive pair of points in the `points` list. The `area` is the absolute value of `shoelace_sum` divided by 2.0. `i` is equal to `len(points) - 2`, and `x1` is the first element of `points[i]`, `y1` is the second element of `points[i]`, while `x2` is the first element of `points[i + 1]`, and `y2` is the second element of `points[i + 1]`.
```python
print(area)
```
**Output:** **area (where area is the area of the polygon formed by the points in the `points` list)**

---

### Example 3  
**Initial State:** `balances` is a list of integers, `A` is the first element of the `balances` list, `B` is the second element of the `balances` list, and `amount` is an integer less than or equal to `A`.
```python
print(f"The amount {amount} is less than or equal to {A}")
```
**Output:** **The amount [amount] is less than or equal to [A] (where `amount` is the value of `amount` and `A` is the first element of the `balances` list)**

---

### **Your Task**

**Initial State:**  
{pre}

```python
{program}
```

Now, please think step by step. Based on the precondition that describes the state of the program, variables, objects, etc., before the code is executed, calculate what will be printed when the `print` statement is executed. Explain the values of the variables, objects, etc., that are printed.

Use natural language to describe the output, ensuring it is easily understandable by a human. Strictly adhere to the format **Output: `what is printed`**.

\end{lstlisting}
\end{tcolorbox}

\clearpage
\flushbottom
\subsection{Return Command Execution Prompt}
\label{return_prompt}
\begin{tcolorbox}[ colback=gray!10, colframe=black, arc=5pt]
\begin{lstlisting}[language=, basicstyle=\ttfamily\footnotesize]
You have been assigned the role of a program executor, responsible for simulating the execution of Python code. You will be provided with an **initial state** and a **Python code snippet**. Your task is to determine what the program **returns** based on the given initial state.
- Avoid describing how the program runs step by step.
- If a variable has a **specific value**, use that value directly.
- If a return statement is executed, **always mention the returned value explicitly** in the output state.
- Adhere to the text format: **Output State: `output state`**.

I am giving you some examples to understand the task better. Then I am giving you your task.
### Example 1  
**Initial State:** `str` is a string, 'str' has 3 or more characters  
```python
return str
```
**Output State:** **The program returns string `str` that has 3 or more characters**

### Example 2  
**Initial State:** `numbers` is an empty list, `total` is the sum of all positive integers that were in the original `numbers` list, `count` is the number of positive integers that were in the original `numbers` list, `average` is equal to `total / count`  
```python
return average
```
**Output State:** **The program returns `average`, which is equal to `total/count`, where `total` is the sum of all positive integers in the original `numbers` list, and `count` is the number of positive integers in the original `numbers` list.**

### Example 3  
**Initial State:** `n` is either 3 or 5  
```python
return n + 1
```
**Output State:** **The program returns 4 or 6**

### Example 4  
**Initial State:** `x` is 1, `y` is greater than 3, `z` is 0  
```python
return y + x
```
**Output State:** **The program returns `y + 1`, where `y` is greater than 3**

### Example 5  
**Initial State:** `count` contains the number of numbers greater than 1 in the list `numbers`, `numbers` is a list of integers, `total` is 0  
```python
return count
```
**Output State:** **The program returns the number of integers greater than 1 in list `numbers`**

### **Your Task**

**Initial State:**  
{pre}

```python
{program}
```

Now, please think step by step: List the impact of the code on the program, check the previous values of the affected variables, and then calculate what the program returns. Any variable or value that is included in the return should be fully described with all available information.

Strictly adhere to the format: **Output State: `the output state you calculate`**, and describe this output state in **natural language easily understandable by humans**.

\end{lstlisting}
\end{tcolorbox}

\subsection{Try-Except Statement Execution Prompt}
\label{try_except_prompt}
\begin{tcolorbox}[ colback=gray!10, colframe=black, arc=5pt]
\begin{lstlisting}[language=, basicstyle=\ttfamily\footnotesize]
You have been assigned the role of a program verifier, responsible for summarizing the state of the function after executing a Python `try` statement. You will be provided with the **final state of the program** after executing the `try` block, along with the **changes in the program after executing one or more `except` blocks** in any situation. Your task is to **combine this information to summarize the program's state after the complete execution of the `try` statement**.

- If a `return` statement is executed, **always include it in the output state**.
- Adhere to the text format: **Output State: `output state`**.

I am giving you some examples to understand the task better. Then I am giving you your task.

---

### Example 1  
**Initial State:** `a` is an integer, `b` is an integer.  
```python
try:
    result = a / b
    return result
except ZeroDivisionError:
    return None
```
**Output state after executing the `try` block:** `a` is an integer, `b` is an integer, `result` is `a` divided by `b`, and the function returns `result`.

**Output state after executing the `except` block:** If a `ZeroDivisionError` occurs (i.e., `b` is 0), the function returns `None`.

**Final Output State:**  
A `ZeroDivisionError` might be triggered at `result = a / b`. If `b` is 0, the function returns `None`. Otherwise, the function returns the value of `a` divided by `b`.  
**Output State:** **`a` and `b` are integers. If `b` is zero, the function returns `None`, otherwise the function returns the value of `a` divided by `b`.**

---

### Example 2  
**Initial State:** `file_path` is a string that's supposed to be a path to a file.  
```python
def read_file(file_path):
    try:
        with open(file_path, 'r') as file:
            data = file.read()
            print("File content successfully read.")
            return data
    except FileNotFoundError:
        print("Error: The file was not found. Please check the file path.")
        return None
    except PermissionError:
        print("Error: You do not have permission to read this file.")
        return None
```
**Output state after executing the `try` block:** `file_path` is a string representing a file path, `data` contains the file content, and the function returns `data`.

**Output state after executing the `except` block(s):**
- If a `FileNotFoundError` occurs, the function prints an error message and returns `None`.
- If a `PermissionError` occurs, the function prints a different error message and returns `None`.

**Final Output State:**  
The program could raise a `FileNotFoundError` if the file is not found at the specified path or a `PermissionError` if the user does not have permission to read the file. If the file is found and the user has permission, the function reads the file content and returns it.  
**Output State:** **`file_path` is a string representing a file path. If the file exists and the user has permission, the function returns the file content; otherwise, it prints an error message and returns `None`.**

\end{lstlisting}
\end{tcolorbox}

\begin{tcolorbox}[ colback=gray!10, colframe=black, arc=5pt]
\begin{lstlisting}[language=, basicstyle=\ttfamily\footnotesize]
### **Your Task**

**Initial State:**  
{pre}

**Code for the try-except block:**
```python
{code}
```

**Output state after executing the `try` block:**  
{try_post}

**Output state after executing the `except` block(s):**  
{except_post}

Now, please think step by step: At which point in the program could such an exception occur? Summarize what the `try-except` statement accomplishes and what the program output state is after execution.

Strictly adhere to the format: **Output State: `the output state you calculate`**, and describe this output state in **natural language easily understandable by humans**.

\end{lstlisting}
\end{tcolorbox}

\subsection{Function Functionality and Return Summary Prompt}
\label{functionality_prompt}
\begin{tcolorbox}[ colback=gray!10, colframe=black, arc=5pt]
\begin{lstlisting}[language=, basicstyle=\ttfamily\footnotesize]
You have been assigned the task of a program verifier, responsible for describing the functionality of a Python function. You will be provided with the **constraints and relationships** between the input parameters, as well as the **resulting output** of the function based on these inputs. Your task is to **organize this information and describe the functionality of the function**.

- Avoid describing how the function operates (e.g., local variable initialization or step-by-step execution).
- Focus only on **what parameters the function accepts** and **what it returns**.
- If there are multiple return points in the function, we will split them into **cases** labeled sequentially (Case 1, Case 2, etc.).
- If one case is fulfilled, **none of the others are executed**.
- Adhere strictly to the format: **Functionality: `functionality description`**.

I am giving you some examples to understand the task better. Then I am giving you your task.


### Example 1  
**Parameter constraints and relationships:** `number` is an integer.  
```python
def func(number):
```
**Output:**  
- Case 1: If `number` is even, the function returns `True`.
- Case 2: If `number` is not even, the function returns `False`.

**Final Answer:**  
The function `func` accepts a parameter `number` and returns `True` if `number` is even. If `number` is not even, it returns `False`.  
**Functionality:** **The function accepts a parameter `number`, returns `True` if `number` is even. If `number` is not even, it returns `False`.**

### Example 2  
**Parameter constraints and relationships:** `age` is an integer.  
```python
def func(age):
```
**Output:**  
- Case 1: If `age` is less than 0, the function returns an error message stating that ages can't be negative.
- Case 2: If `age` is between 0 and 18, the function returns "minor".
- Case 3: Otherwise, the function returns "adult".

**Final Answer:**  
The function `func` accepts a parameter `age`. If `age` is below 0, the function returns an error message. If `age` is between 0 and 18, it returns "minor". Otherwise, it returns "adult".  
**Functionality:** **The function accepts an integer `age`, returns an error if `age` is negative, "minor" if `age` is between 0 and 18, and "adult" otherwise.**

### **Your Task**

**Parameter constraints:**  
{pre}

```python
{head}
```

**Output:**  
{body_post}

Now, please think step by step: What are the parameters the function accepts, and what does it return?

Strictly adhere to the format: **Functionality: `the functionality you calculate`**, and describe this functionality in **natural language easily understandable by humans**.

\end{lstlisting}
\end{tcolorbox}

\subsection{If Condition Precondition Extraction Prompt}
\label{if_pre_prompt}
\begin{tcolorbox}[ colback=gray!10, colframe=black, arc=5pt]
\begin{lstlisting}[language=, basicstyle=\ttfamily\footnotesize]
You are assigned the role of a program verifier, responsible for finding the **postcondition** of an `if` statement based on its condition. You will be given:
- A **precondition**, describing the initial state of the program variables before entering the `if` statement.
- An **if condition**, which determines whether the program enters the `if` block.

Your task is to determine the **postcondition**, which describes the state of the program variables after entering the `if` block. The postcondition must extend the precondition by ensuring that the `if` condition is true. This description should include both the values of the variables and the relationships between them.

- **Do not explain how the program runs**; only focus on variable values and relationships.
- Ensure that the postcondition retains all valid information from the precondition while incorporating the truth of the `if` condition.
- Follow the format: **Postcondition: `calculated postcondition`**.

I am giving you some examples to understand the task better. Then I am giving you your task.

---

### Example 1  
**Precondition:** `str` is a string  
**If condition:**  
```python
if len(str) < 3:
```
**Postcondition:** **`str` is a string, and the length of `str` is less than 3**

---

### Example 2  
**Precondition:** `n` can have any value  
**If condition:**  
```python
if isinstance(n, int):
```
**Postcondition:** **`n` is an integer of any value**

---

### Example 3  
**Precondition:** `x` is a positive integer  
**If condition:**  
```python
if x < 2:
```
**Postcondition:** **`x` is a positive integer less than 2**

---

### Example 4  
**Precondition:** `m` is an integer. If `m` is higher than 10, then `n` equals `-m`. `a` is a list of integers.  
**If condition:**  
```python
if n < 0:
```
**Postcondition:** **`m` is an integer, if `m` is higher than 10, `n` equals `-m`. `a` is a list of integers. The current value of `n` is less than 0**

\end{lstlisting}
\end{tcolorbox}

\begin{tcolorbox}[ colback=gray!10, colframe=black, arc=5pt]
\begin{lstlisting}[language=, basicstyle=\ttfamily\footnotesize]
### Example 5  
**Precondition:** `x` is an integer, `a` is a list of integers.  
**If condition:**  
```python
if a[0] != 0:
```
**Postcondition:** **`x` is an integer, `a` is a list of integers. The first element of `a` is not 0**

### **Your Task**

**Precondition:**  
{precondition}

**If condition:**  
```python
{program_fragment}
```

Your task is to complete the **postcondition**. Ensure that:
- All valid information from the **precondition** is retained.
- The **if condition** is now assumed to be **true**.
- If variables have values related to previous conditions, put that early in the postcondition and state the **current** value of the variable clearly.
- The final state of the program **after entering the if block** is fully described.

Strictly adhere to the format: **Postcondition: `the postcondition you calculate`**, and describe this postcondition in **natural language easily understandable by humans**.

\end{lstlisting}
\end{tcolorbox}

\subsection{Else Condition Precondition Extraction Prompt}
\label{else_pre_prompt}
\begin{tcolorbox}[ colback=gray!10, colframe=black, arc=5pt]
\begin{lstlisting}[language=, basicstyle=\ttfamily\footnotesize]
You are assigned the role of a program verifier, responsible for finding the **postcondition** of an `else` statement based on the condition of the corresponding `if` statement. You will be given:
- A **precondition**, describing the initial state of the program variables before evaluating the `if` condition.
- An **if condition**, which determines whether the program enters the `if` block. **In this case, we do not enter the `if` block but instead follow the `else` block**.

Your task is to determine the **postcondition**, which describes the state of the program variables after entering the `else` block. The postcondition must extend the precondition by ensuring that the `if` condition is **false**. This description should include both the values of the variables and their relationships.

- **Do not explain how the program runs**; only focus on variable values and relationships.
- Ensure that the postcondition retains all valid information from the precondition while incorporating the negation of the `if` condition.
- Follow the format: **Postcondition: `calculated postcondition`**.

I am giving you some examples to understand the task better. Then I am giving you your task.

---

### Example 1  
**Precondition:** `str` is a string  
**If condition:**  
```python
if len(str) < 3:
```
**Postcondition:** **`str` is a string, and the length of `str` is greater than or equal to 3**

---

### Example 2  
**Precondition:** `n` can have any value  
**If condition:**  
```python
if isinstance(n, int):
```
**Postcondition:** **`n` can have any value except an integer**

---

### Example 3  
**Precondition:** `x` is a positive integer  
**If condition:**  
```python
if x < 2:
```
**Postcondition:** **`x` is a positive integer greater than or equal to 2**

---

### Example 4  
**Precondition:** `m` is an integer, `n` is an integer, `a` is a list of integers.  
**If condition:**  
```python
if n <= 0:
```
**Postcondition:** **`m` and `n` are integers, `n` is greater than 0, `a` is a list of integers**

\end{lstlisting}
\end{tcolorbox}
\begin{tcolorbox}[ colback=gray!10, colframe=black, arc=5pt]
\begin{lstlisting}[language=, basicstyle=\ttfamily\footnotesize]
### Example 5  
**Precondition:** `x` is an integer, `a` is a list of integers.  
**If condition:**  
```python
if a[0] != 0:
```
**Postcondition:** **`x` is an integer, `a` is a list of integers. The first element of `a` is 0**

### **Your Task**

**Precondition:**  
{precondition}

**If condition:**  
```python
{program_fragment}
```

Your task is to complete the **postcondition**. Ensure that:
- All valid information from the **precondition** is retained.
- The **if condition** is now assumed to be **false**.
- If variables have values related to previous conditions, put that early in the postcondition and state the **current** value of the variable clearly.
- The final state of the program **after entering the else block** is fully described.

Strictly adhere to the format: **Postcondition: `the postcondition you calculate`**, and describe this postcondition in **natural language easily understandable by humans**.

\end{lstlisting}
\end{tcolorbox}

\subsection{If-Else Statement Postcondition Extraction Prompt}
\label{if_else_prompt}
\begin{tcolorbox}[ colback=gray!10, colframe=black, arc=5pt]
\begin{lstlisting}[language=, basicstyle=\ttfamily\footnotesize]
You are assigned the role of a program verifier, responsible for completing the **overall postcondition** of Hoare triples for `if` statements based on the postconditions of both the `if` and `else` parts. You will be given:
- A **precondition**, describing the initial state of the program variables before the execution of the program fragment.
- A **program fragment**, which contains an `if` condition and possibly an `else` block.
- The **postcondition of the `if` part**, describing the state after entering the `if` block.
- The **postcondition of the `else` part**, describing the state after entering the `else` block (if an `else` exists).

Your task is to **combine the postconditions of both the `if` and `else` parts** (if an `else` exists), taking into account the `if` condition, to derive the **overall postcondition of the entire `if-else` block**.

- **Do not explain how the program runs**; only focus on variable values and relationships.
- Ensure that the postcondition retains all valid information from the precondition while incorporating both branches of the `if-else` logic.
- Summarize the `if` statement in a **coherent paragraph**, rather than discussing it in separate sections.
- Follow the format: **Postcondition: `calculated postcondition`**.

I am giving you some examples to understand the task better. Then I am giving you your task.

---

### Example 1  
**Precondition:** `str` is a string  
**Program Fragment:**  
```python
if len(str) < 3:
```
**If part:** `str` is a string with length less than 3, the function returns `None`  
**Else part:** There is no `else` part in the code  

**Postcondition:** **`str` is a string. If the length of `str` is less than 3, the function returns `None`.**

---

### Example 2  
**Precondition:** `n` can have any value  
**Program Fragment:**  
```python
if isinstance(n, int):
    
else:
```
**If part:** `n` is an integer of any value, and the function returns `n`  
**Else part:** `n` can have any value except an integer, and the function returns `int(n)`  

**Postcondition:** **If `n` is an integer, the function returns `n` itself. Otherwise, the function returns `int(n)`.**

---

### Example 3  
**Precondition:** `x` is a positive integer  
**Program Fragment:**  
```python
if x < 2:
    
else:
```
**If part:** `x` is a positive integer less than 2, and the function returns `False`  
**Else part:** `x` is a positive integer greater than or equal to 2, and the function returns `True`  

**Postcondition:** **`x` is a positive integer. If `x` is less than 2, the function returns `False`. Otherwise, the function returns `True`.**


\end{lstlisting}
\end{tcolorbox}

\begin{tcolorbox}[ colback=gray!10, colframe=black, arc=5pt]
\begin{lstlisting}[language=, basicstyle=\ttfamily\footnotesize]
### Example 4  
**Precondition:** `m` is an integer, `n` is an integer  
**Program Fragment:**  
```python
if n < 0:
    
else:
```
**If part:** The integer `n` was originally negative, `n` is updated to its negation, and `m` is increased by 1  
**Else part:** If `n` is 0, the function returns `m`. Otherwise, `n` is decreased by 13, and `m` is increased by 1  

**Postcondition:** **`m` and `n` are integers. If `n < 0`, `m` is increased by 1 and `n` is negated. If `n == 0`, the function returns `m`. Otherwise, `n` is decreased by 13 and `m` is increased by 1.**

---

### Example 5  
**Precondition:** `x` is an integer, `y` is zero  
**Program Fragment:**  
```python
if x > 0:
```
**If part:** `x` is a positive integer. If `x > 10`, `y` is set to twice the value of `x`. Otherwise, `y` is set to `x + 5`.  
**Else part:** There is no `else` part in the code  

**Postcondition:** **`x` is an integer. If `x > 10`, `y` is set to twice the value of `x`. If `x` is greater than 0 but less than 10, `y` is set to `x + 5`.**

---

### **Your Task**

**Precondition:**  
{precondition}

**Program Fragment:**  
```python
{program_fragment}
```

**If part:**  
{postconditions_if}

**Else part:**  
{postconditions_else}

Your task is to **complete the overall postcondition of the entire if-else block**. Summarize all cases and describe the final state of the program **after executing the if-else statement**.

Strictly adhere to the format: **Postcondition: `the postcondition you calculate`**, and describe this postcondition in **natural language easily understandable by humans**.

\end{lstlisting}
\end{tcolorbox}

\subsection{While Loop First Execution Necessary Condition Prompt}
\label{while_first_prompt}
\begin{tcolorbox}[ colback=gray!10, colframe=black, arc=5pt]
\begin{lstlisting}[language=, basicstyle=\ttfamily\footnotesize]
You have been assigned the task of a program verifier, responsible for modifying the **program state** so that the first iteration of the `while` loop can proceed. You will be provided with the **program state right before the loop**, which you need to modify, and the `while` loop statement.

- If the loop is a `while True` loop or if the loop **can certainly execute at least once**, simply repeat the program state right before the loop.
- **Do not make any assumptions** beyond the provided information.
- Only modify the states of **objects in the loop condition** that affect whether the loop executes.
- **Follow the text format:** **State: `state`**.

I am giving you some examples to understand the task better. Then I am giving you your task.

### Example 1  
**State right before the while loop:** `total` is 10, `i` is 0, `n` is an integer.  
**While loop:**  
```python
while i < n:
    # the loop body is omitted
```
Now, please think step by step: Which states need to be adjusted for the loop to execute the first time? **Only the states of objects in the loop head can be adjusted.**

**Example Answer:**  
The variables in the loop condition are `i` and `n`, so we can only adjust them. The loop executes while `i < n`. Right before the loop, `i` is 0 and `n` is an integer. Since `n` being an integer does not ensure the loop executes, it needs to be modified to `n` is greater than 0.

**State:** **`total` is 10, `i` is 0, `n` must be greater than 0**

### Example 2  
**State right before the while loop:** `total` is 0, `students` is 2 less than its initial value.  
**While loop:**  
```python
while students >= 1:
    # the loop body is omitted
```
Now, please think step by step: Which states need to be adjusted for the loop to execute the first time? **Only the states of objects in the loop head can be adjusted.**

**Example Answer:**  
The only variable in the loop condition is `students`. The loop executes while `students >= 1`. Right before the loop, `students` is **2 less than its initial value**, so for the loop to execute at least once, the initial value of `students` must have been at least 3, and its current value must be greater than or equal to 1.

**State:** **`total` is 0, `students` is 2 less than its initial value, and `students` must currently be greater than or equal to 1**

### **Your Task**

**State right before the while loop:**  
{post}

```python
{loop_head}
    # the loop body is omitted
```

Now, please think step by step: Which states need to be adjusted for the loop to execute the first time? **Only modify the states of objects in the loop head.*
Strictly adhere to the format: **State: `the state you calculate`**, and describe this state in **natural language easily understandable by humans**.

\end{lstlisting}
\end{tcolorbox}

\subsection{While Loop Prerequisite for Next Execution Prompt}
\label{while_next_prompt}
\begin{tcolorbox}[ colback=gray!10, colframe=black, arc=5pt]
\begin{lstlisting}[language=, basicstyle=\ttfamily\footnotesize]
You have been assigned the task of a program verifier, responsible for modifying the **program state** so that the next iteration of the `while` loop can proceed. You will be provided with the **program state after the previous iteration**, which you need to modify, and the `while` loop statement.

- If the loop is a `while True` loop or if the loop **can certainly execute one more time**, simply repeat the program state at the end of the previous iteration.
- **Do not make any assumptions** beyond the provided information.
- Only modify the states of **objects in the loop condition** that affect whether the loop executes.
- **Follow the text format:** **State: `state`**.

I am giving you some examples to understand the task better. Then I am giving you your task.

### Example 1  
**State at the end of the previous iteration:** `total` is 10, `i` is 4, `n` is greater than 3.  
**While loop:**  
```python
while i < n:
    # the loop body is omitted
```
Now, please think step by step: Which states need to be adjusted for the loop to execute one more time? **Only the states of objects in the loop head can be adjusted.**

**Example Answer:**  
The variables in the loop condition are `i` and `n`, so we can only adjust them. The loop executes while `i < n`. At the end of the last iteration, `i` is 4, `n` is greater than 3. `n` being greater than 3 does not ensure another execution, so it needs to be modified to `n` is greater than 4.

**State:** **`total` is 10, `i` is 4, `n` must be greater than 4**


### Example 2  
**State at the end of the previous iteration:** `total` is 0, `students` is 3 less than its initial value.  
**While loop:**  
```python
while students >= 1:
    # the loop body is omitted
```
Now, please think step by step: Which states need to be adjusted for the loop to execute one more time? **Only the states of objects in the loop head can be adjusted.**

**Example Answer:**  
The only variable in the loop condition is `students`. The loop executes while `students >= 1`. At the end of the last iteration, `students` is **3 less than its initial value**, so for the loop to execute again, the initial value of `students` must have been at least 4, and its current value must be greater than or equal to 1.

**State:** **`total` is 0, `students` is 3 less than its initial value, and `students` must currently be greater than or equal to 1**

### **Your Task**

**State at the end of the previous iteration:**  
{post}

```python
{loop_head}
    # the loop body is omitted
```

Now, please think step by step: Which states need to be adjusted for the loop to execute one more time? **Only modify the states of objects in the loop head.**
Strictly adhere to the format: **State: `the state you calculate`**, and describe this state in **natural language easily understandable by humans**.

\end{lstlisting}
\end{tcolorbox}

\subsection{For Loop First Execution Necessary Condition Prompt}
\label{for_first_prompt}

\begin{tcolorbox}[ colback=gray!10, colframe=black, arc=5pt]
\begin{lstlisting}[language=, basicstyle=\ttfamily\footnotesize]
You have been assigned the task of a program verifier, responsible for understanding the **program state at the start of the `for` loop**. You will be provided with the **program state before the first execution of the `for` loop**, which you need to modify, and the `for` loop statement.

- **Do not make any assumptions** beyond the provided information.
- Only modify the states of **objects in the loop condition** that affect whether the loop executes.
- **Follow the text format:** **State: `state`**.

I am giving you some examples to understand the task better. Then I am giving you your task.

### Example 1  
**State before the `for` loop:** `total` is 10.  
**For loop:**  
```python
for i in range(n):
    # the loop body is omitted
```
Now, please think step by step: Which states need to be adjusted for the loop to execute? **Only the states of objects in the loop head can be adjusted.**

**Example Answer:**  
The only variables in the loop head are `i` and `n`, so we can only adjust those. The loop executes while `n` is at least 1. Since `total` being 10 does not ensure that the loop will execute, `n` needs to be adjusted to be greater than 0, and `i` is initialized to 0.

**State:** **`total` is 10, `i` is 0, `n` must be greater than 0**


### Example 2  
**State before the loop starts:** `total` is 0, `students_list` is a list of students.  
**For loop:**  
```python
for index, student in enumerate(students_list):
    # the loop body is omitted
```
Now, please think step by step: Which states need to be adjusted for the loop to execute? **Only the states of objects in the loop head can be adjusted.**

**Example Answer:**  
The only objects in the loop head are `index`, `student`, and `students_list`, so we can only adjust those. The loop executes if `students_list` has at least 1 student. Since the `total` value does not impact execution, we focus on ensuring that `students_list` contains at least 1 student. The `index` starts at 0, and `student` is set to the first student in the list.

**State:** **`total` is 0, `students_list` is a list of students that must have at least 1 student, `student` is the first student in the list, `index` is 0**

### **Your Task**

**State before the loop starts:**  
{post}

```python
{loop_head}
    # the loop body is omitted
```

Now, please think step by step: Which states need to be adjusted for the loop to execute? **Only modify the states of objects in the loop head.**

Strictly adhere to the format: **State: `the state you calculate`**, and describe this state in **natural language easily understandable by humans**.

\end{lstlisting}
\end{tcolorbox}

\subsection{For Loop Prerequisite for Next Execution Prompt}
\label{for_next_prompt}
\begin{tcolorbox}[ colback=gray!10, colframe=black, arc=5pt]
\begin{lstlisting}[language=, basicstyle=\ttfamily\footnotesize]
You have been assigned the task of a program verifier, responsible for understanding the **program state at the start of the next iteration of a `for` loop**. You will be provided with the **program state after the previous iteration**, which you need to modify, and the `for` loop statement.
- **Do not make any assumptions** beyond the provided information.
- Only modify the states of **objects in the loop condition** that affect whether the loop executes again.
- **Follow the text format:** **State: `state`**.

I am giving you some examples to understand the task better. Then I am giving you your task.

### Example 1  
**State at the end of the previous iteration:** `total` is 10, `i` is 3, `n` must be greater than 3.  
**For loop:**  
```python
for i in range(n):
    # the loop body is omitted
```
Now, please think step by step: Which states need to be adjusted at the start of the next iteration of the loop? **Only the states of objects in the loop head can be adjusted.**

**Example Answer:**  
The only variables in the loop head are `i` and `n`, so we can only adjust those. The loop executes while `i` is less than `n`. At the end of the last iteration, `i` is 3, `n` is greater than 3. Since `i` is incremented by 1 in each iteration, for the loop to execute again, `i` must be 4 and `n` must be greater than 4.

**State:** **`total` is 10, `i` is 4, `n` must be greater than 4**

### Example 2  
**State at the end of the previous iteration:** `total` is 0, `students_list` is a list of students that must have at least 2 students, `student` is the second student in the list, `index` is 1.  
**For loop:**  
```python
for index, student in enumerate(students_list):
    # the loop body is omitted
```
Now, please think step by step: Which states need to be adjusted for the loop to execute one more time? **Only the states of objects in the loop head can be adjusted.**

**Example Answer:**  
The only objects in the loop head are `index`, `student`, and `students_list`, so we can only adjust those. The loop executes if `students_list` has at least 3 students. At the end of the last iteration, `students_list` has at least 2 students, `student` is the second student in the list, and `index` is 1. So, for the loop to execute one more time, `students_list` must have at least 3 students, `index` must be 2, and `student` must be the third student in the list.

**State:** **`total` is 0, `students_list` is a list of students that must have at least 3 students, `student` is the third student in the list, `index` is 2**

### **Your Task**

**State at the end of the previous iteration:**  
{post}

```python
{loop_head}
    # the loop body is omitted
```

Now, please think step by step: Which states need to be adjusted for the loop to execute one more time? **Only modify the states of objects in the loop head.**
Strictly adhere to the format: **State: `the state you calculate`**, and describe this state in **natural language easily understandable by humans**.
\end{lstlisting}
\end{tcolorbox}

\subsection{Total While Loop Execution with Unrolling Prompt}
\label{total_while_prompt}
\begin{tcolorbox}[ colback=gray!10, colframe=black, arc=5pt]
\begin{lstlisting}[language=, basicstyle=\ttfamily\footnotesize]
Given a Python loop, an initial execution state, and the output states after the first {times} iterations of the loop, determine the output state after all the executions of the loop have finished.

You must adhere to the text format: Output State: **output state.**

Initial State: {pre}
Code of the loop:
{loop_code}

The output state after the loop executes the first {times} times includes what needed to be true for the loop to execute at least that number of times:
{loop_unrolled}

What is the ouput state after the loop executes all the iterations? Change the values of only the variables in the loop head and body. The state of the other variables in the precondition that are not affected by the loop head and body must remain unchanged.
In your response strictly use the format: Output State: **the output state you calculate.**, and describe this output state in Natural language easily understandable by humans.

\end{lstlisting}
\end{tcolorbox}

\subsection{Total While Loop Execution Without Unrolling Prompt}
\label{while_NO_UNROLL_PROMPT}
\begin{tcolorbox}[ colback=gray!10, colframe=black, arc=5pt]
\begin{lstlisting}[language=, basicstyle=\ttfamily\footnotesize]
Given a Python loop and an initial execution state, determine the output state after all the executions of the loop have finished.

You must adhere to the text format: Output State: **output state.**

Initial State: {pre}
Code of the loop:
{loop_code}

What is the output state after the loop executes all the iterations? Change the values of only the variables in the loop head and body. The state of the other variables in the precondition that are not affected by the loop head and body must remain unchanged.
In your response strictly use the format: Output State: **the output state you calculate.**, and describe this output state in Natural language easily understandable by humans.

\end{lstlisting}
\end{tcolorbox}

\subsection{Total For Loop Execution with Unrolling Prompt}
\label{total_for_prompt}
\begin{tcolorbox}[ colback=gray!10, colframe=black, arc=5pt]
\begin{lstlisting}[language=, basicstyle=\ttfamily\footnotesize]
Given a Python loop, an initial execution state, and the output states after the first {times} iterations of the loop, determine the output state after all the executions of the loop have finished.

You must adhere to the text format: Output State: **output state.**

Initial State: {pre}
Code of the loop:
{loop_code}

The output state after the loop executes the first {times} of times includes what needed to be true for the loop to execute at least that number of times:
{loop_unrolled}

What is the ouput state after the loop executes all the iterations? Change the values of only the variables in the loop head and body. The state of the other variables in the precondition that are not affected by the loop head and body must remain unchanged.
In your response strictly use the format: Output State: **the output state you calculate.**, and describe this output state in Natural language easily understandable by humans.

\end{lstlisting}
\end{tcolorbox}

\subsection{Total For Loop Execution Without Unrolling Prompt}
\label{for_NO_UNROLL_PROMPT}
\begin{tcolorbox}[ colback=gray!10, colframe=black, arc=5pt]
\begin{lstlisting}[language=, basicstyle=\ttfamily\footnotesize]
Given a Python loop, and an initial execution state, determine the output state after all the executions of the loop have finished. 

You must adhere to the text format: Output State: **output state.**

Initial State: {pre}
Code of the loop:
{loop_code}

What is the ouput state after the loop executes all the iterations? Change the values of only the variables in the loop head and body. The state of the other variables in the precondition that are not affected by the loop head and body must remain unchanged.
The output state must be in a similar format as the initial execution state. Describe this output state in Natural language easily understandable by humans. In your response strictly use the format: Output State: **the output state you calculate.**.

\end{lstlisting}
\end{tcolorbox}

\end{document}